\documentclass[aps,showpacs,twocolumn]{revtex4}

\usepackage[dvips]{graphics,graphicx}
\usepackage{amsmath}
\usepackage{amssymb}                 
\usepackage{amscd}                   
\usepackage{wasysym}
\usepackage[ansinew]{inputenc}
\usepackage[T1]{fontenc}
\usepackage{ae,aecompl}

\newcommand{\ri}{{ \rm i }}
\newcommand{\re}{{ \rm e }}
\newcommand{\rd}{{ \rm d }}

\newcommand{\be}{\begin{equation}}
\newcommand{\ee}{\end{equation}}

\newcommand{\cn}{{\rm cn}}
\newcommand{\sn}{{\rm sn}}
\newcommand{\dn}{{\rm dn}}

\renewcommand{\vec}[1]{\mathbf{#1}}

\usepackage{color}
\definecolor{blau}{rgb}{0,0,1}
\definecolor{gruen}{rgb}{0,1,0}
\definecolor{rot}{rgb}{1,0,0}
\definecolor{magenta}{rgb}{1,0,1}

\begin{document}
\bibliographystyle{apsrev}
\title{An analytical study of resonant transport of Bose--Einstein condensates}
\author{K. Rapedius, D. Witthaut, and H. J. Korsch}
\email{korsch@physik.uni-kl.de}
\affiliation{Technische Universit{\"a}t Kaiserslautern, FB Physik,
                            D-67653 Kaiserslautern, Germany}
\date{\today }

\begin{abstract}

\noindent
We study the stationary nonlinear Schr\"odinger equation,
or Gross--Pitaevskii equation, for a one--dimensional finite square well potential.
By neglecting the mean--field interaction outside the potential well it is possible to discuss the transport properties of the system analytically in terms of ingoing and outgoing waves. Resonances and bound states are obtained analytically. The transmitted flux shows a bistable behaviour. Novel crossing scenarios of eigenstates similar to beak--to--beak structures are observed for a repulsive mean--field interaction. It is proven that resonances transform to bound states due to an attractive nonlinearity and vice versa for a repulsive nonlinearity, and the critical nonlinearity for the transformation is calculated analytically. The bound state wavefunctions of the system satisfy an oscillation theorem as in the case of linear quantum mechanics. Furthermore, the implications of the eigenstates on the dymamics of the system are discussed.

\end{abstract}
\maketitle

\section{Introduction}

Due to the experimental progress in the field of Bose-Einstein condensates (BECs), e.g.~atom--chip experiments, achieved in the past few years \cite{Hans01,Folm00,Ott01}, it is now possible to study the influence of an interatomic interaction on transport. At low temperatures, BECs can be described  by the nonlinear Schr\"odinger equation (NLSE) or Gross-- Pitaevskii Equation (GPE)
\be
  \ri \hbar \frac{\partial \psi(\vec{r},t)}{\partial t}  =
  \left( -\frac{\hbar^2}{2 m}  \nabla^2 + V(\vec{r})
  + g | \psi(\vec{r}, t) |^2 \right)  \psi(\vec{r},t)
   \label{GPGl}
\ee
in a mean-field approach \cite{Pita03,Park98,Dalf99,Legg01}. Another important application of the NLSE is the propagation of electromagnetic waves in nonlinear media (see, e.g., \cite{Dodd82}, ch. 8).
In the case of vanishing interaction ($g=0$), Eq.~(\ref{GPGl}) reduces to the linear Schr\"odinger equation of single particle quantum mechanics.

The treatment of transport within this mean-field theory reveals new interesting phenomena which arise from the nonlinearity of the equation. For example a bistable behaviour of the flux transmitted through a double barrier potential has been found \cite{Paul05}. For the same system transitions from resonances to bound states can occur due to the mean-field interaction \cite{Schl04,Mois03}. However, previous studies of nonlinear transport and nonlinear resonances usually rely strongly on numerical solutions, such as complex scaling. There are only few model systems of the NLSE which have been studied analytically and most of these treatments concentrate on bound states. Among these few examples are the infinite square well \cite{Carr00a,Carr00b,Dago00}, the finite square well \cite{Carr01,Lebo01}, the potential step \cite{Seam05a}, delta-potentials \cite{Witt05,Haki97} and the delta-comb \cite{Seam05b, Witt05b} as an example of a periodic potential.
Therefore analytical solutions of the NLSE for other simple model potentials are of fundamental interest.

In the present paper we solve the time-independent NLSE for a simple one-dimensional square-well potential in order to analytically indicate and discuss the nonlinear phenomena described above. To unambiguously define ingoing and outgoing waves and thus a transmission coefficient $|T|^2$, we neglect the mean-field-interaction $g|\psi(x)|^2$ outside the potential well.
Since the one-dimensional mean--field coupling constant $g$ is inversely proportional to the square of the system's radial extension $a_\bot$ this can be achieved in an atom--chip experiment by a weaker radial confinement of the condensate outside the potential well \cite{Pita03,Lebo01,Olsh98}.
Especially for bound states or very stable resonances our approximation is well justified since in these cases the condensate density $|\psi(x)|^2$ is much higher inside the potential well than outside.

The paper is organized as follows: In section \ref{LSE_Scat}, we briefly review some basic results for the scattering states of the linear Schr\"odinger equation with a square well potential which are treated in various textbooks on quantum mechanics \cite{Schw98,Mess69,Nolt92}. In section \ref{NLSE_Scat} we will then analyze stationary scattering states of the NLSE with a finite square-well-potential. In section \ref{LSE_bs} we again have a look at the linear system with respect to bound states in order to prepare for discussing the transition from resonances to bound states for an attractive (section \ref{AT_bs}) and a repulsive mean--field interaction (section \ref{Rep_bs}). In addition to our discussion in terms of stationary states we have a brief look at the dynamics of the system in section \ref{dyn}, where we solve the time--dependent NLSE numerically.

\section{Linear Schr\"odinger equation: Scattering}
\label{LSE_Scat}
The scattering states of the linear time-independent Schr\"odinger Equation
\be
  \left( -\frac{\hbar^2}{2m} \frac{d^2}{dx^2} + V(x) \right) \psi(x)=E \psi(x)
  \label{schrod}
\ee
for a particle with mass $m$ where the potential is given by
\be
   V(x)=\left\{
                    \begin{array}{cc}
                     V_0,   &  |x| \le a\\
                     0,   &     |x|>a
                    \end{array}
              \right.
          \label{square}
\ee
with $V_0<0$ are obtained for positive particle energies $E>0$.
We assume an incoming flux from $x=-\infty$ and make the ansatz
\be
 \psi(x)= \left\{
                    \begin{array}{cc}
                     A \,{\rm e}^{{\rm i}kx}+B \, {\rm e}^{-{\rm i}kx}   &    x<-a \\
                     F \, {\rm e}^{{\rm i}qx} +G \, {\rm e}^{-{\rm i}qx} &  |x| \le a\\
                     C \, \re^{{\rm i}kx}              &     x>a \, ,
                    \end{array}
              \right.
\ee
where the wavenumbers inside ($|x| \le a$) and outside ($|x|>a$) the potential well are given by
\be
   q=\sqrt{2m(E-V_0)}/\hbar .
\ee
and
\be
   k=\sqrt{2mE}/\hbar ,
   \label{well_lin}
\ee
respectively.
The amplitudes of transmission and reflexion are then determined by
\be
   T:=C/A \quad \text{and} \quad R:=B/A .
\ee
They satisfy the relation $|R|^2+|T|^2=1$ as the Schr\"odinger equation is flux-preserving. We have to make the wavefunction and its derivative continuous at $x=\pm a$.
Together with the normalisation condition $A=1$ these boundary conditions determine the wavefunction of the system. By eliminating $F$ and $G$ we obtain the amplitude of transmission
\be
   T(E)=\frac{1}{\cos(2qa)} \frac{\exp(-2\ri ka)}{1-(\ri/2)[(q/k)+(k/q)]\tan^2(qa)}
   \label{Tlin}
\ee
and thus the transmission probability
\be
 |T(E)|^2 = \left(1+\frac{1}{4}\left(\frac{q}{k}-\frac{k}{q}\right)^2 \sin^2(2qa) \right)^{-1} \, .
 \label{Tqlin}
\ee
\begin{figure}[htb]
\centering
\includegraphics[width=8cm,  angle=0]{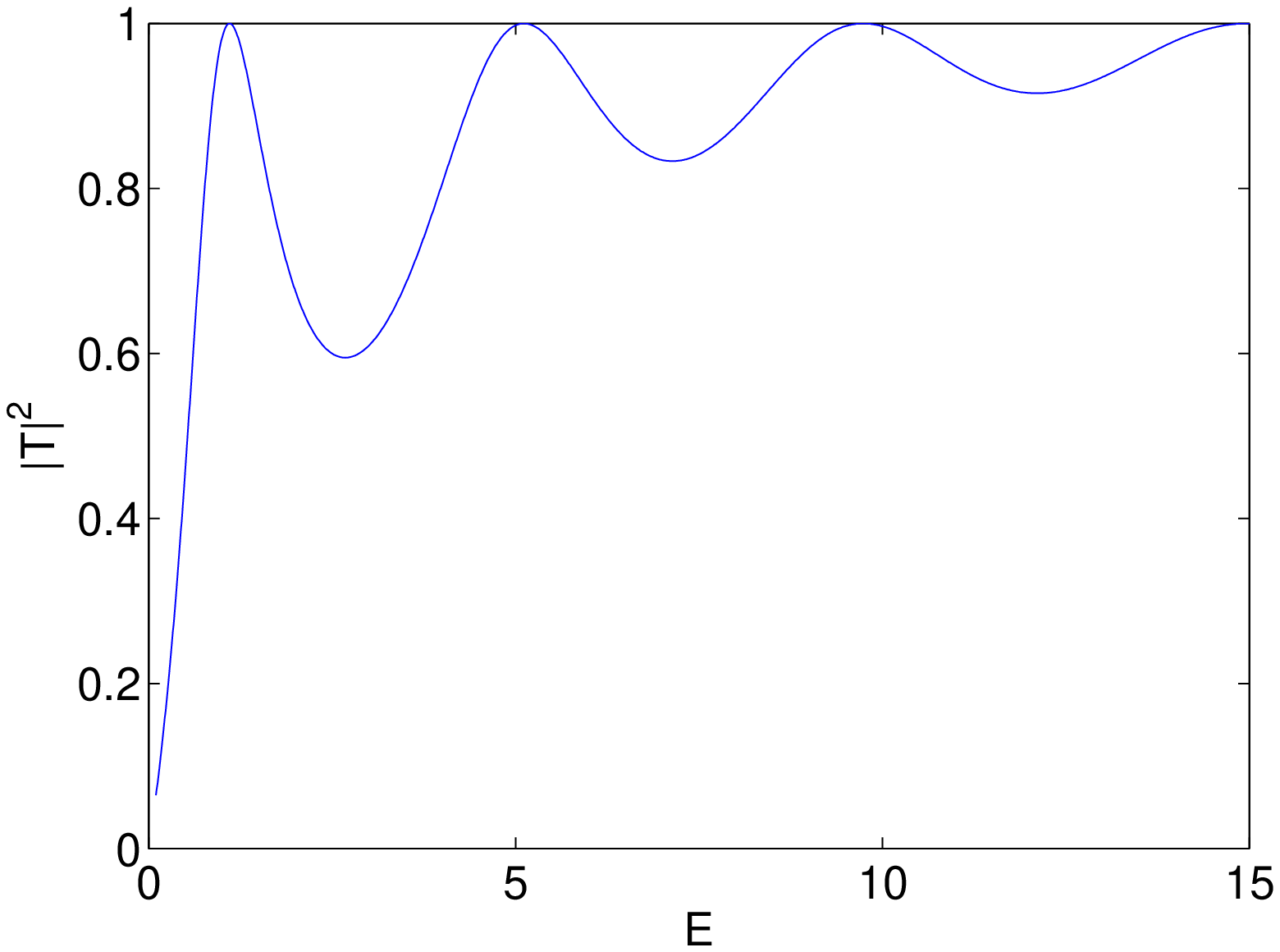}
\hspace{5mm}
\includegraphics[width=8cm,  angle=0]{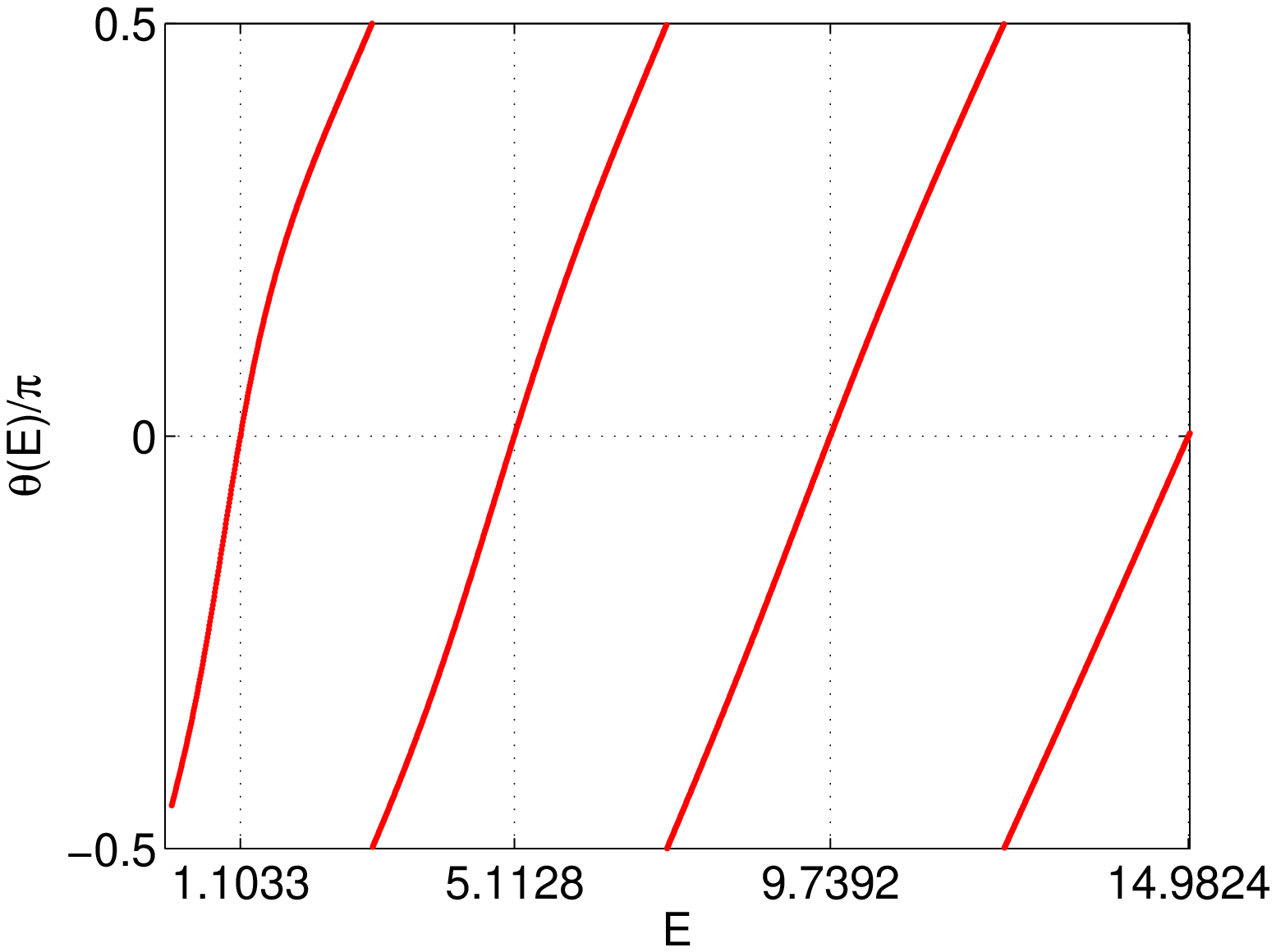}
\caption{\label{fig-T_lin} {(Color online) Transmission probability (left) and scattering phase (right) for the potential $V_0=-10$ and $a=2$ (for $\hbar=m=1$).}}
\end{figure}
We define a resonance as a maximum of the transmission coefficient where the potential well is fully transparent, that is $|T(E)|^2=1$. Then Eq.~(\ref{Tqlin}) implies the resonance condition
\be
   2qa=n \pi .
\ee
The resonance energies are then given by
\be
   E_n=V_0+ \frac{\hbar^2 q^2}{2m}=V_0+ \frac{\hbar^2 \pi^2}{8ma^2}n^2 ,
   \label{ERlin}
\ee
where $n$ is an integer number and sufficiently large to make $E_n>0$.
A Taylor expansion of Eq.~(\ref{Tlin}) in terms of $E-E_n$ yields
\be
   |T(E)|^2 \approx \frac{(\Gamma_n/2)^2}{(E-E_n)^2 +(\Gamma_n/2)^2} \, ,
   \label{Lorentz}
\ee
where $\Gamma_n$ is given by
\be
   \Gamma_n=\frac{2 \sqrt{E_n}(E_n-V_0)}{2E_n-V_0} .
   \label{Gamma_n}
\ee
This means that the transmission coefficient can be approximated by a Lorentzian with width $\Gamma_n$ in the vicinity of the resonances. We define the scattering phase $\theta(E)$ by
\be
   T(E)=|T(E)|\re^{\ri \theta(E)-2\ri ka}
\ee
and obtain
\be
   \tan \theta(E)=\frac{1}{2} \left( \frac{q}{k} +\frac{k}{q} \right) \tan(2qa) .
\ee
In the vicinity of a resonance this reduces to
\be
   \theta(E) \approx \mbox{arctan} \left(\frac{2}{\Gamma_n}(E-E_n) \right) .
\ee
Note that the scattering phase is zero for $E=E_n$.
The bound states of this system will be discussed in section \ref{LSE_bs}.

\section{NLSE: Scattering}
\label{NLSE_Scat}

Now we consider the same square well potential with an additional mean-field term $g|\psi(x)|^2$ inside the potential well so that for $|x|>a$ the wavefunction must satisfy the free linear Schr\"odinger equation
\be
   \left( -\frac{\hbar^2}{2m} \frac{d^2}{dx^2} -\mu \right) \psi(x) = 0
   \label{sq_aussen}
\ee
and for $|x| \le a$ the time--independent NLSE
\be
    \left( -\frac{\hbar^2}{2m} \frac{d^2}{dx^2} + g|\psi(x)|^2 +V_0 -\mu \right) \psi(x) =
    0\,
    \label{sq_innen}
\ee
with a constant potential $V_0<0$, where $\mu$ is the chemical potential.

For an incoming flux from $x=-\infty$ we now make the ansatz
\be
 \psi(x)= \left\{
                    \begin{array}{cl}
                     A \, {\rm e}^{{\rm i}kx}+B \, {\rm e}^{-{\rm i}kx}   &    x<-a \\
                     f(x)                                           &   |x| \le a\\
                     C \, \re^{{\rm i}kx}                              &     x>a
                    \end{array}
              \right.
\ee
with the wavenumber (\ref{well_lin}).
Inside the potential we assume an intrinsically complex solution
\be
          f(x)=\sqrt{S(x)} \re^{\ri \Phi(x)}
	  \label{f(x)}
\ee
where $S(x)$ and $\Phi(x)$ are real functions. As already shown by Carr, Clark and Reinhard \cite{Carr00a,Carr00b}  a complex solution of the NLSE with a constant potential is given by the equations
\be
   \Phi'(x)=\frac{\alpha}{S(x)}
\ee
and
\be
   S(x)=\varepsilon+\varphi \, \dn^2\left(\varrho x+\delta|p\right)
   \label{S(x)}
\ee
in terms of the Jacobi elliptic function $\dn$ with the real period $2K(p)$, where $K(p)$ is the complete elliptic integral of the first kind and $p$ is the elliptic parameter \cite{Abra72,Lawd89,Bowm61}.
The real parameters $\alpha, p, \varepsilon, \varphi, \varrho$ and $\delta$ have to satisfy the relations
\be
 \varphi = - \frac{\hbar^2}{gm}\varrho^2
 \label{I}
\ee
\be
 \mu =V_0+ \frac{3}{2} g \varepsilon -\frac{\hbar^2}{2m}(2-p)\varrho^2
 \label{II}
\ee
\be
 -\frac{\hbar^2}{2m}\left(\varrho^2 \varphi \varepsilon (p-1) -\alpha^2\right)
+ g \varepsilon^3 - (\mu-V_0) \varepsilon^2 =0  \, .
  \label{III}
\ee
Again, the wavefunction and its derivative must be continuous at $x=\pm a$.
At $x=+a$ we obtain the equations
\be
   f(a)=C \, \re^{\ri ka}, \quad   f'(a)=\ri k  \, C \, \re^{\ri ka} \, .
   \label{5.36}
\ee
By inserting Eq.~(\ref{f(x)}) we arrive at
\be
   \frac{1}{2}S'(a)+\ri \alpha = \ri k \, S(a) \, .
   \label{Sp_alpha}
\ee
Since we are considering positive chemical potentials $\mu>0$, the wavenumber $k$ is real so that we obtain the conditions
\be
   S'(a)=0
   \label{Sp_a_0}
\ee
and
\be
   \alpha=k S(a) .
   \label{alp_kSa}
\ee
At $x=-a$ we obtain
\be
   A \, \re^{-\ri ka} + B \, \re^{\ri ka} = f(-a)
   \label{RB1}
\ee
and
\be
   \ri k \left( A \, \re^{-\ri ka} + B \, \re^{\ri ka} \right)=f'(-a).
\ee
With the equations (\ref{f(x)}) and (\ref{alp_kSa}) these two conditions can be written as
\be
   \sqrt{S(-a)}A \, \re^{-\ri ka}=\frac{1}{2}\left[ S(-a)+S(a)-\ri \frac{S'(-a)}{2k} \right] \re^{\ri \Phi(-a)}
   \label{Ae_neu}
\ee
\be
   \sqrt{S(-a)}B \, \re^{+\ri ka}=\frac{1}{2}\left[ S(-a)-S(a)+\ri \frac{S'(-a)}{2k} \right]\re^{\ri \Phi(-a)}
   \label{Be_neu} .
\ee
As in the linear case, we define the amplitudes of reflection and transmission as $R:=B/A$ and $T:=C/A$.
Equations (\ref{Ae_neu}) and (\ref{Be_neu}) then lead to the reflection probability
\be
   |R|^2=\frac{[S(-a)-S(a)]^2+S'(-a)^2/(4k^2)}{[S(-a)+S(a)]^2+S'(-a)^2/(4k^2)}
   \label{Rq}
\ee
and the transmission probability
\be
   |T|^2=\frac{4S(a)S(-a)}{[S(-a)+S(a)]^2+S'(-a)^2/(4k^2)}=\frac{S(a)}{|A|^2} \, .
   \label{Tq}
\ee
It can easily be verified that $|R|^2+|T|^2=1$. The scattering phase $\theta(\mu)$, defined by $T(\mu)=|T(\mu)| {\rm exp}(\ri \theta(\mu)-2\ri ka)$, can be expressed as
\begin{eqnarray}
   \theta(\mu)&=& \arg(C)-\arg(A)+2ka \\
         &=& \int_{-a }^{+a}\frac{kS(a)}{S(x)} \rd x+\arctan \frac{S'(-a)}{2k \left(S(-a)+S(a) \right)} . \nonumber
   \label{SPhase}
\end{eqnarray}

The condition (\ref{Sp_a_0}) can be further exploited by rewriting it as
\be
 -2 \varphi \varrho p \ \mbox{dn}(u|p) \, \mbox{sn}(u|p) \, \mbox{cn}(u|p) = 0 \, ,
 \label{5.57}
\ee
with the abbreviation $u=\varrho a+\delta$.
As the $\dn$-function is always nonzero, the solutions of Eq.~(\ref{5.57}) are given by the zeros of either of the two functions $\cn(u|p)$ or $\sn(u|p)$.

\begin{description}
\item [Case 1]: $\sn(\varrho a+\delta|p)=0$
\be
       \Leftrightarrow \varrho a +\delta = (2j+1)K(p)
    \ee
    \be
       \Rightarrow \delta=(2j+1)K(p)-\varrho a \, .
    \ee
The variable $j$ denotes an integer number. This means
\be
   \mbox{dn}^2\left(\varrho a +\delta|p\right)=\mbox{dn}^2(2jK(p)|p)=1
\ee
or
\be
   S(a)=\varepsilon+\varphi.
   \label{Sa_ep_phi}
\ee

\item [Case 2]: $\cn(\varrho a+\delta|p)=0$

In analogy to case 1 we get
\be
   \delta=(2j+1)K(p)-\varrho a
\ee
and
\be
   S(a)=\varepsilon+\varphi(1-p).
   \label{Sa_ep_phi_2}
\ee
\end{description}

For given values of the chemical potential $\mu$ and the incoming amplitude $|A|$, the parameters $p, \varepsilon,$ and $\varphi=-\hbar^2\varrho^2/gm$ are completely determined by the equations (\ref{II}) and (\ref{III}) and the absolute square of Eq.~(\ref{Ae_neu}),
\be
   16 k^2 |A|^2S(-a)=4k^2[S(-a)+S(a)]^2+S'(-a)^2 .
   \label{IV}
\ee

The positions of the resonances are defined by $|T|^2=1$ or $|R|^2=0$. With Eq.~(\ref{Rq}) this is equivalent to the conditions
\be
   S'(-a)=0
   \label{Sp_-a_0}
\ee
and
\be
   S(-a)=S(a)
   \label{Sa_S-a} \, .
\ee

Now we consider these conditions for the two cases mentioned above.

\begin{description}
\item [Case 1]: $\sn(\varrho a+\delta|p)=0$ or  $\delta=(2j+1)K(p)-\varrho a$

The condition (\ref{Sp_-a_0}) now reads
\be \text{ } \text{ } \text{ } \text{ } \text{ }
     S'(-a)=-2 \varphi \varrho p  \, \mbox{cn}(u|p)\, \mbox{sn}(u|p)\, \mbox{dn}(u|p) = 0
\ee
with $u=-\varrho a+\delta$.
The solutions of Eq.~(\ref{Sp_-a_0}) are again given by the zeros of either the $\sn$-function or the $\cn$-function. One can show easily that only the zeros of the $\sn$-function are compatible with the condition (\ref{Sa_S-a}).
Thus we obtain
\be
   -2\varrho a +2jK(p)= \ell K(p)
\ee
where $\ell$ and $j$ are integer numbers. The resonance positions are then given by
\be
     \varrho=\frac{K(p)}{a}n \, ,
     \label{rho_Ka}
\ee
where the integer number $n$ is defined as $n=\ell-j$.

\item [Case 2]: $\cn(\varrho a+\delta|p)=0$ or $\delta=2jK(p)-\varrho a$

 In analogy to case 1 one can show that the resonances are also determined by Eq.~(\ref{rho_Ka}).

\end{description}

In the case of a resonance there is no reflected wave ($B=0$) so that the equations (\ref{RB1}), (\ref{Sa_ep_phi}), (\ref{Sa_ep_phi_2}) and (\ref{Sa_S-a}) lead to
\be
   |A|^2 = S(-a) = S(a)    =  \left\{ \begin{array}{cc}
                                                    \varepsilon+\varphi        &    \text{case 1} \\
                                                    \varepsilon+\varphi (1-p)  &    \text{case 2}\\
                                                    \end{array}
                                             \right.
\ee
or
\be
  \varepsilon (p) =\left\{ \begin{array}{cc}
                            |A|^2-\varphi         &    \text{case 1} \\
                            |A|^2-\varphi (1-p)   &    \text{case 2} \\
                            \end{array}
                     \right. \, .
             \label{ep_p}
\ee
With the help of the equations (\ref{ep_p}), (\ref{I}) and (\ref{II}) we obtain a formula for the chemical potential of a resonance,
\be
  \mu _R(p) = V_0+\frac{3}{2}g|A|^2+\frac{\hbar^2}{2m}\frac{K^2(p)}{a^2} \, n^2 \cdot
                                                         \left\{ \begin{array}{cc}
                                                                  (1+p)    &   \text{case 1}  \\
                                                                  (1-2p)   &   \text{case 2}  \, .
                                                                  \end{array}
                                                         \right.
                                                      \label{muR}
\ee
The chemical potential depends on the parameter $p$ and the integer number $n$ which must be sufficiently large to make $\mu_R \ge 0$.
As in the linear case (cf. Eq. (\ref{ERlin})) there is a term which grows quadratically in $n$. This term can be viewed as an effective increase of the chemical potential in comparison to the energy of the linear system in case 1 and respectively as a decrease in case 2. However, it turns out (cf. below) that the energy shift due to nonlinear interaction is mainly determined by the term $\frac{3}{2}g|A|^2$ which is a direct consequence of the effective increase or decrease of the potential well by the mean-field term $g|\psi(x)|^2$.
As we will show below, we have $p \rightarrow 0$ if $g \rightarrow 0$ so that $K(p) \rightarrow \pi/2$ and Eq.~(\ref{muR}) reduces to the equivalent equation (\ref{ERlin}) for the resonance energies of the linear system.

\begin{figure}[htb]
\centering
\includegraphics[width=8cm,  angle=0]{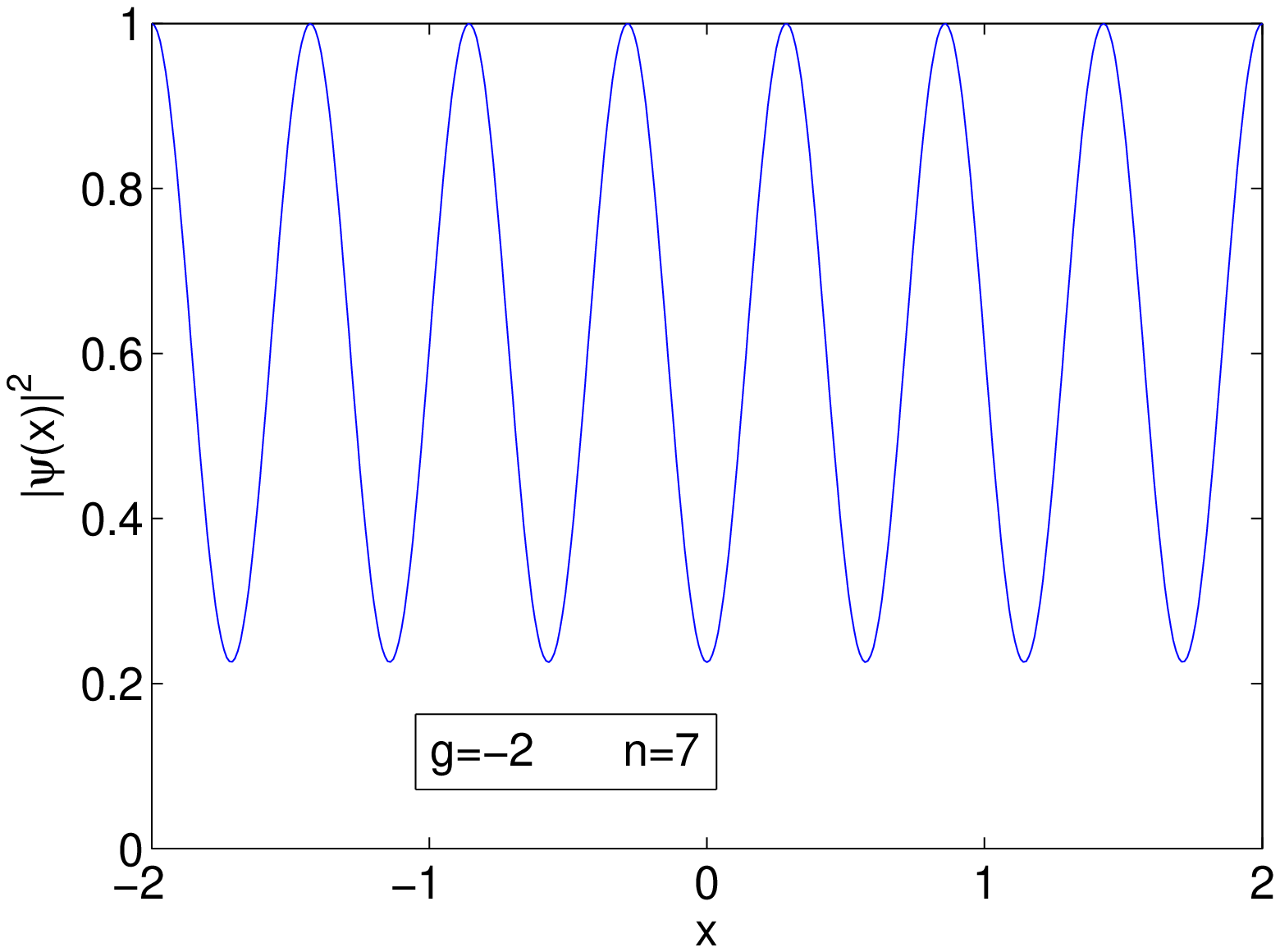}
\hspace{5mm}
\includegraphics[width=8cm,  angle=0]{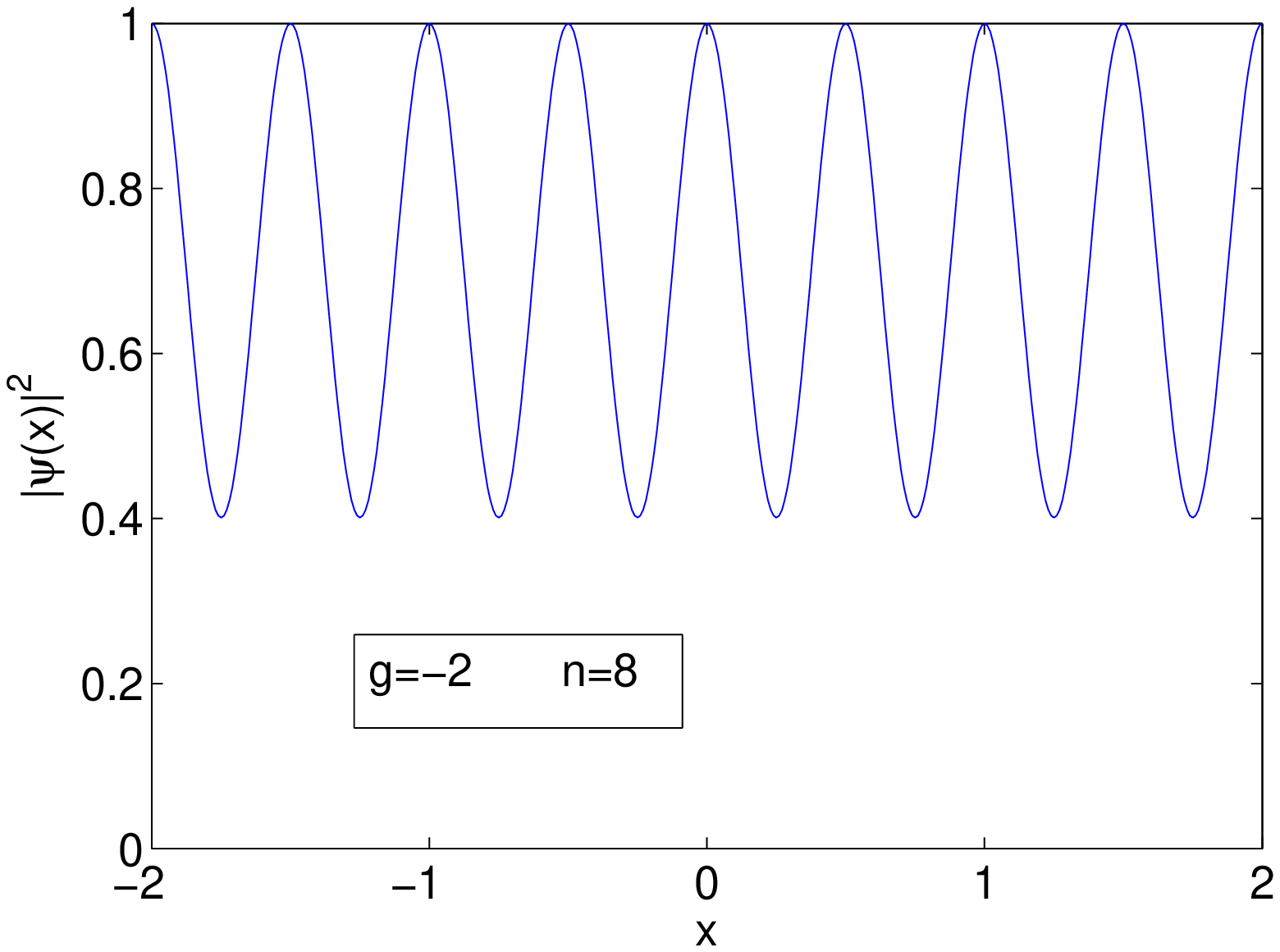}
\caption{\label{fig-WFQ-} {(Color online) Squared modulus of the wavefunctions of the two most stable resonances of the potential $V_0=-10$, $a=2$ for the attractive nonlinearity $g=-2$. The normalization $|A|=1$ has been used.}}
\end{figure}

\begin{figure}[htb]
\centering
\includegraphics[width=8cm,  angle=0]{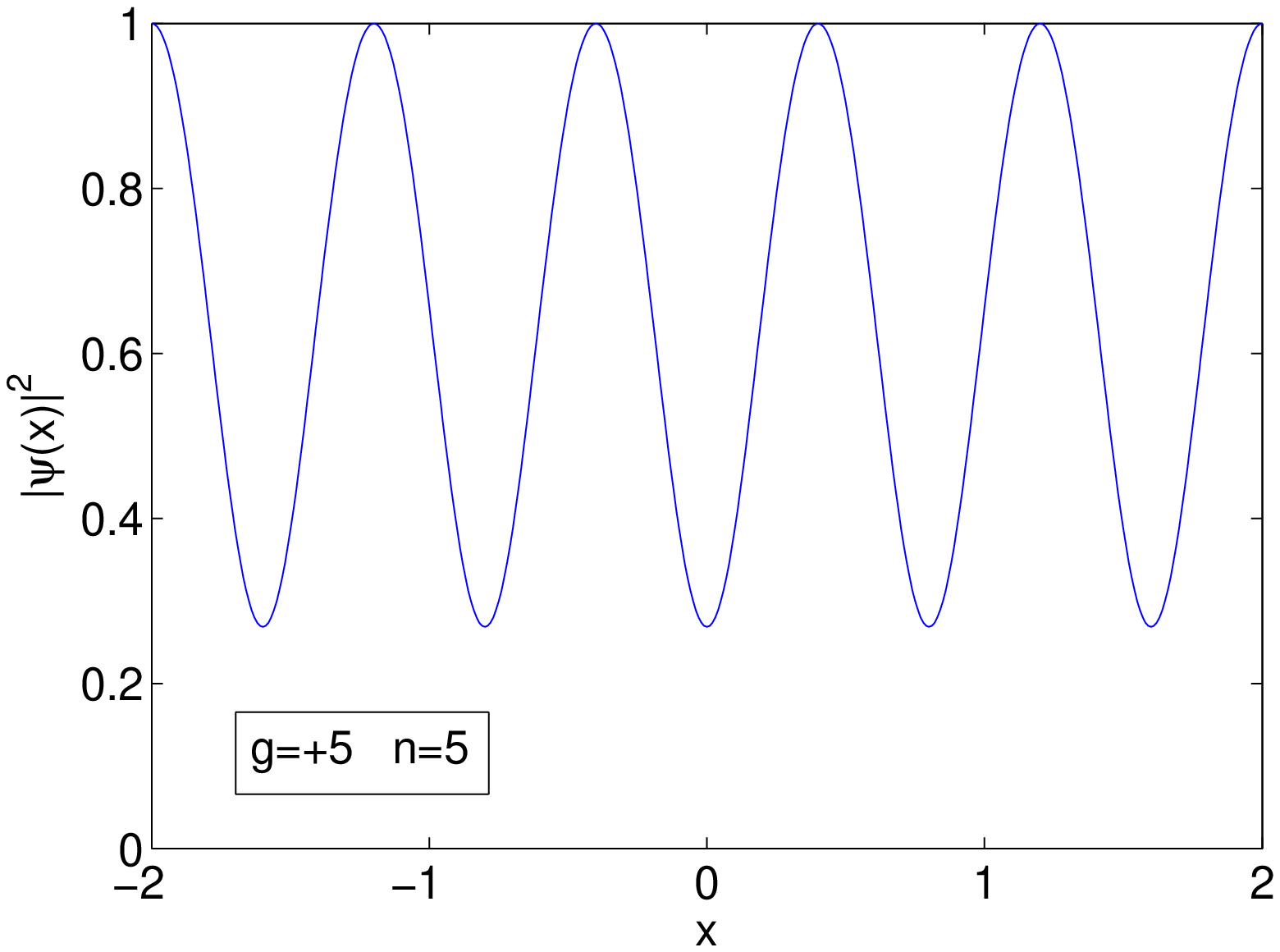}
\hspace{5mm}
\includegraphics[width=8cm,  angle=0]{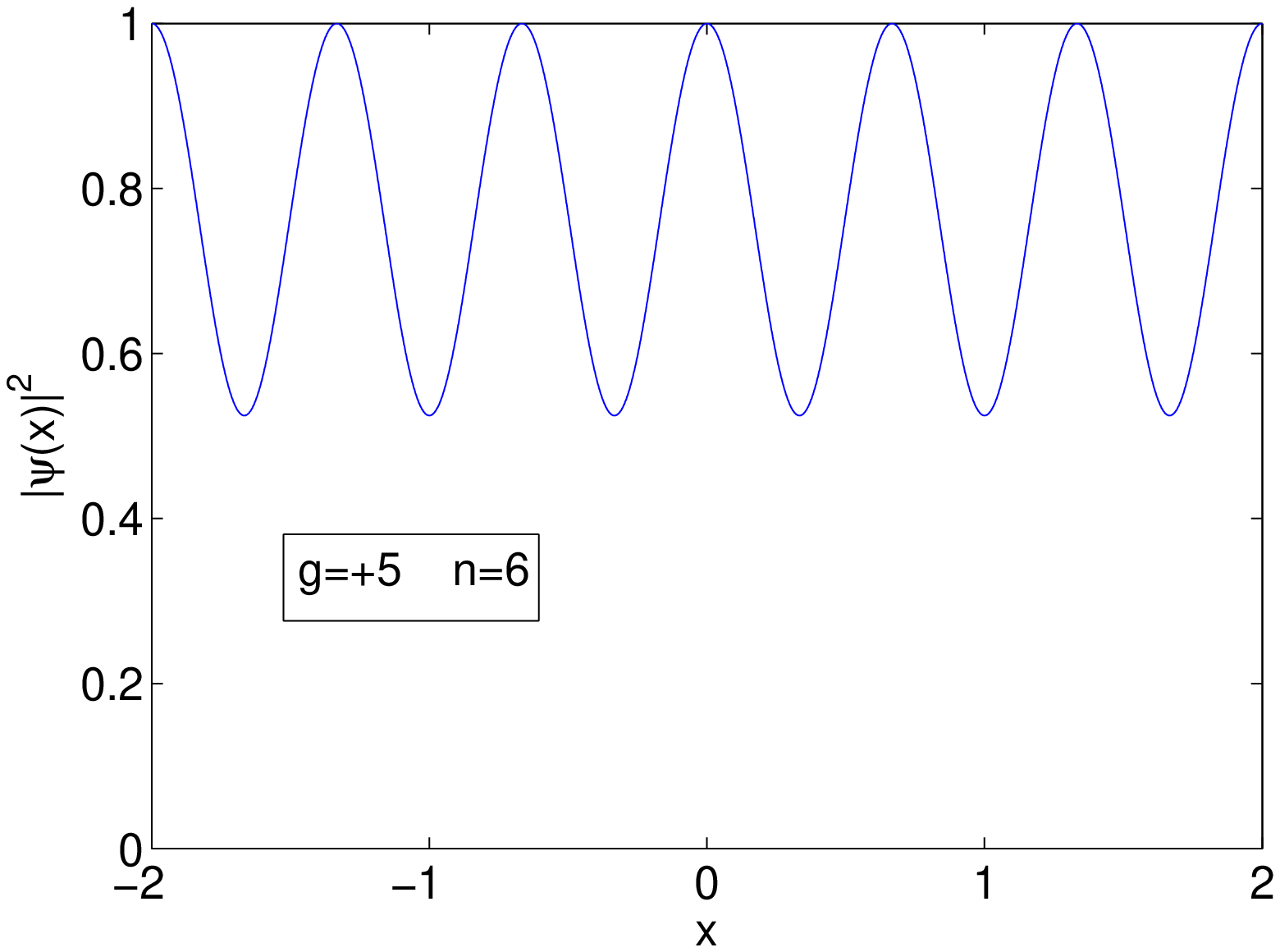}
\caption{\label{fig-WFQ+} {(Color online) Squared modulus of the wavefunctions of the two most stable resonances of the potential $V_0=-12$, $a=4$ for the repulsive nonlinearity $g=+5$. The normalization $|A|=1$ has been used.}}
\end{figure}

The elliptic parameter $p$ can be determined by inserting the expressions (\ref{muR}) and (\ref{ep_p}) into Eq.~(\ref{III}).
We obtain:

\begin{description}
\item [Case 1:]
 \be
       pK^4(p)=g|A|^2(V_0+g|A|^2) \frac{2m^2a^4}{\hbar^4 n^4}
       \label{pK4_1} \, .
    \ee

For $g=0$ this equation can only be solved by $p=0$. The inequality $pK^4(p) \ge 0$ implies
\be
       g|A|^2(V_0+g|A|^2) \ge 0 .
    \ee
Since $V_0$ is negative this means either $g \le 0$ or $g|A|^2 \ge |V_0|$.
For small values of $p$ the left hand side of (\ref{pK4_1}) can be approximated by
\be
       pK^4(p)=\left(\frac{\pi}{2}\right)^4(p+p^2)+\mathcal{O}(p^3)
    \ee
and we obtain
\be \text{ } \text{ } \text{ }
       p\approx-\frac{1}{2}+\sqrt{\frac{1}{4}+\frac{2g|A|^2(V_0+g|A|^2)m^2a^4}{(\pi/2)^4 \hbar^4 n^4} }.
       \label{p_ap1}
    \ee

\item [Case 2:]

 \be \text{ } \text{ } \text{ } \text{ } \text{ }
       p(1-p)K^4(p)=-g|A|^2(V_0+g|A|^2) \frac{2m^2a^4}{\hbar^4 n^4}.
       \label{pK4_2}
    \ee

For $g=0$, this equation is solved by $p=0$. The solution $p=1$ is incompatible with Eq.~(\ref{muR}) because it leads to a negative chemical potential $\mu_R$.
From $p(1-p)K^4(p) \le0$ we obtain the condition
$0 \le g|A|^2 \le |V_0|$.
For small values of $p$, the left hand side of (\ref{pK4_2}) can be approximated by
 \be
       p(1-p)K^4(p)=\left(\frac{\pi}{2}\right)^4p+\mathcal{O}(p^3)
  \ee
and we get
    \be
       p\approx-\frac{2g|A|^2(V_0+g|A|^2)m^2a^4}{(\pi/2)^4 \hbar^4 n^4}.
       \label{p_ap2}
    \ee
\end{description}
Case 2 is valid for $0 \le g \le |V_0|/|A|^2$ and case 1 is valid for any other value of $g$.
Obviously the parameter $p$ and the chemical potential $\mu_R$ only depend on the product $g|A|^2$. Without loss of generality we henceforth assume $|A|^2=1$ in all figures and numerical calculations.
The squared modulus of the resonance wavefunctions inside the potential well $|x| \le a$ in dependence of $p$ and $n$ is given by
\be
  |\psi_R(x)|^2 = |A|^2 -\frac{\hbar^2 \varrho^2}{g} \left[\mbox{dn}^2\left(\varrho x-nK(p)|p\right)-1\right]
  \label{S_R}
\ee
   in case 1 and
\begin{eqnarray}
  |\psi_R(x)|^2 = |A|^2 &-&\frac{\hbar^2\varrho^2}{g} \left[\mbox{dn}^2\left(\varrho x+(1-n)K(p)|p\right) \right. \nonumber\\
                    &-&\left. 1+p\right]
\end{eqnarray}
in case 2 with $\varrho=nK(p)/a$.

\begin{figure}[htb]
\centering
\includegraphics[width=8cm,  angle=0]{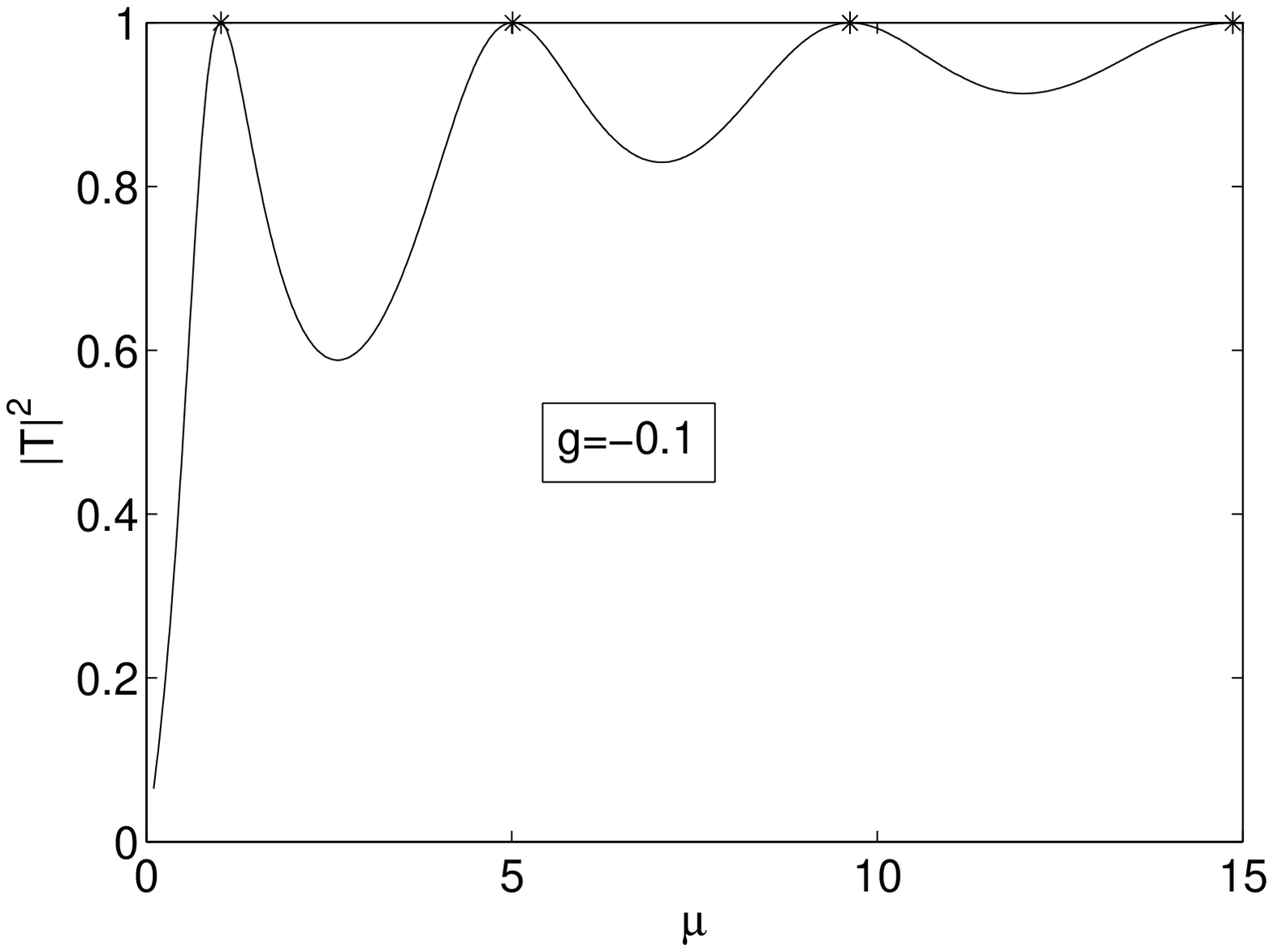}

\includegraphics[width=8cm,  angle=0]{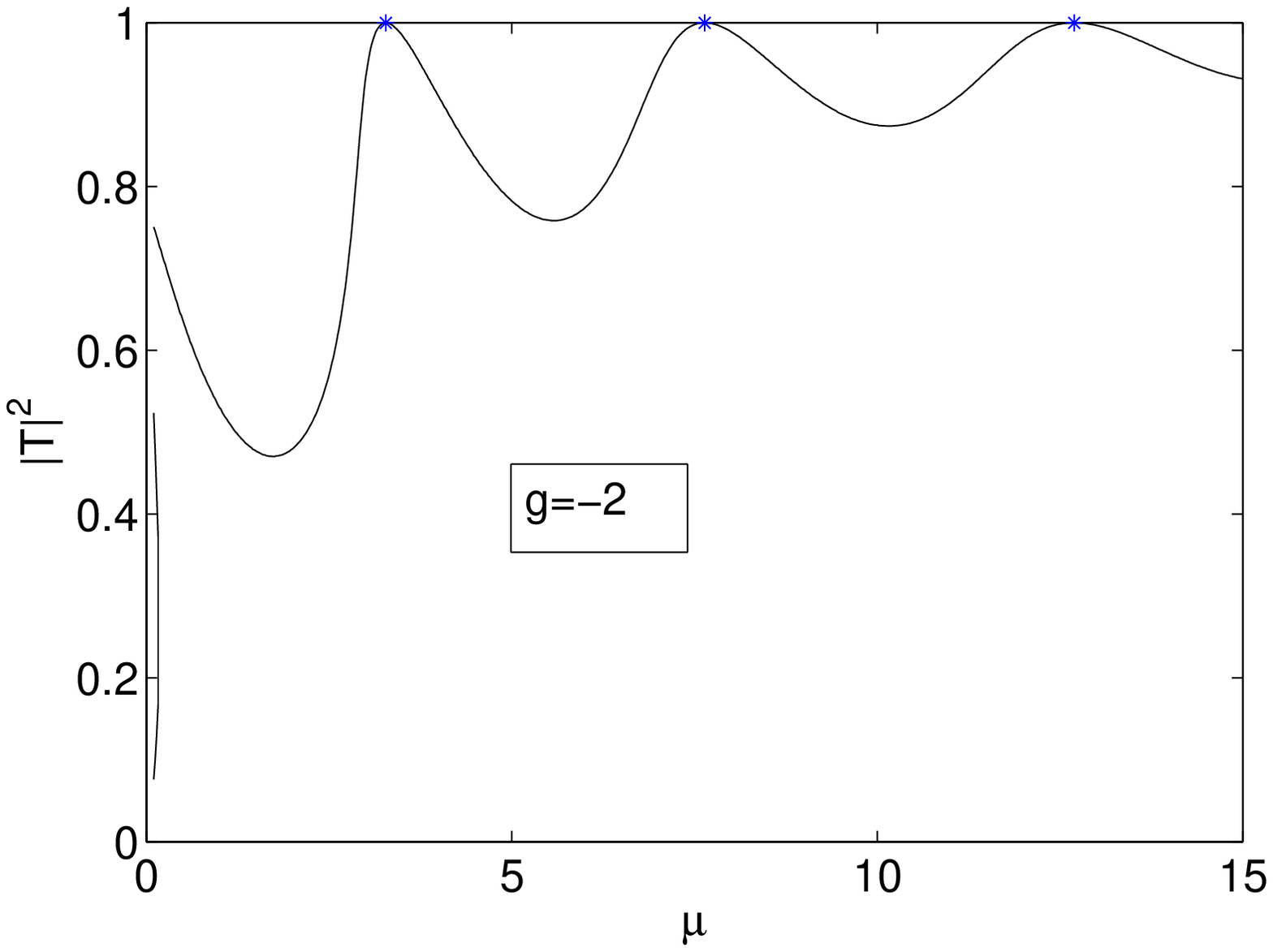}



\caption{\label{fig-T_g-} {(Color online) Transmission coefficients for different
attractive nonlinearities for the square well potential
$V_0=-10$ , $a=2$.
The positions of the resonances, calculated with the formula
(\ref{muR}), are marked by \textcolor{blau}{'*'}.
}}
\end{figure}

\begin{figure}[htb]
\centering


\includegraphics[width=8cm,  angle=0]{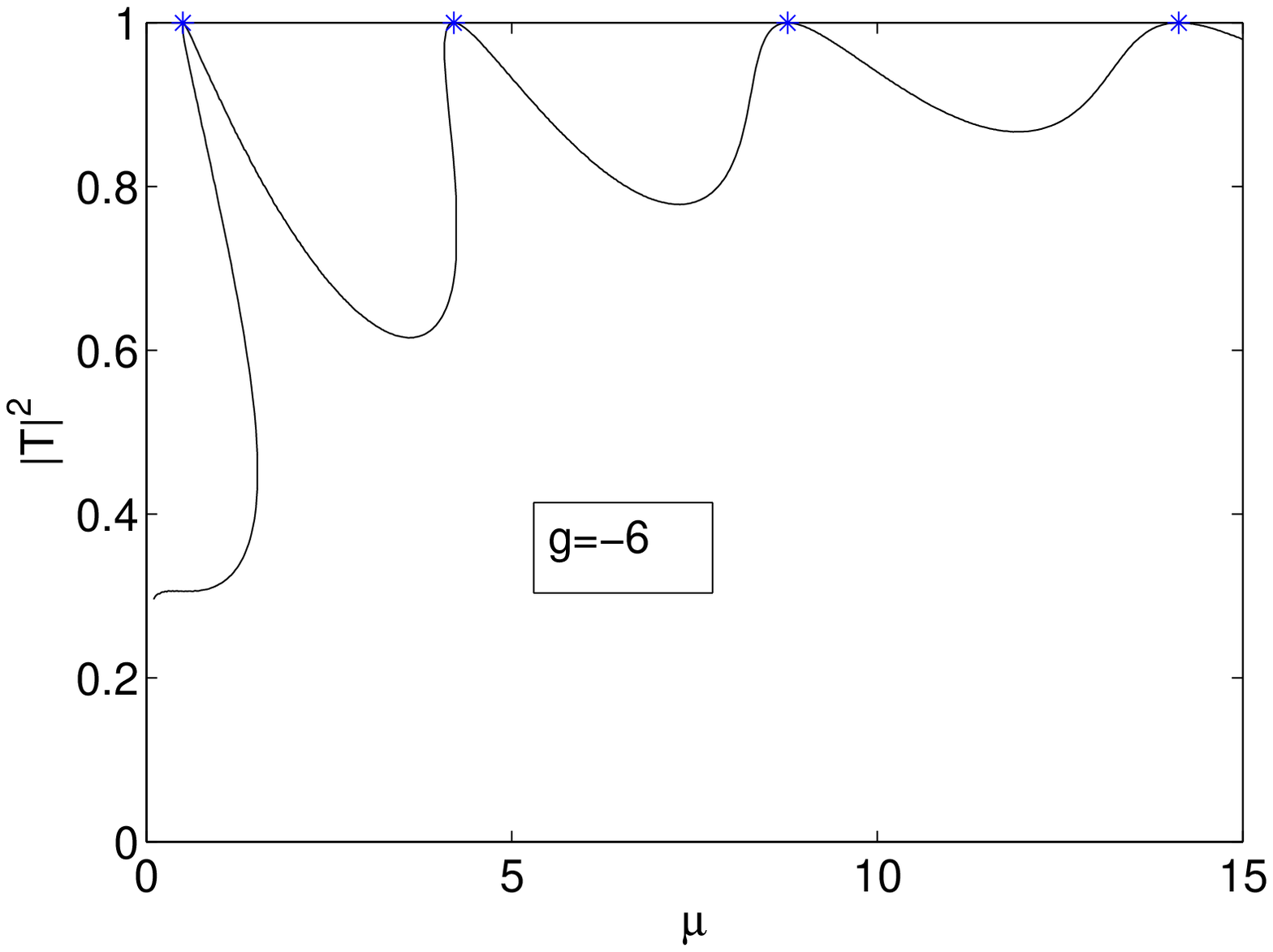}

\includegraphics[width=8cm,  angle=0]{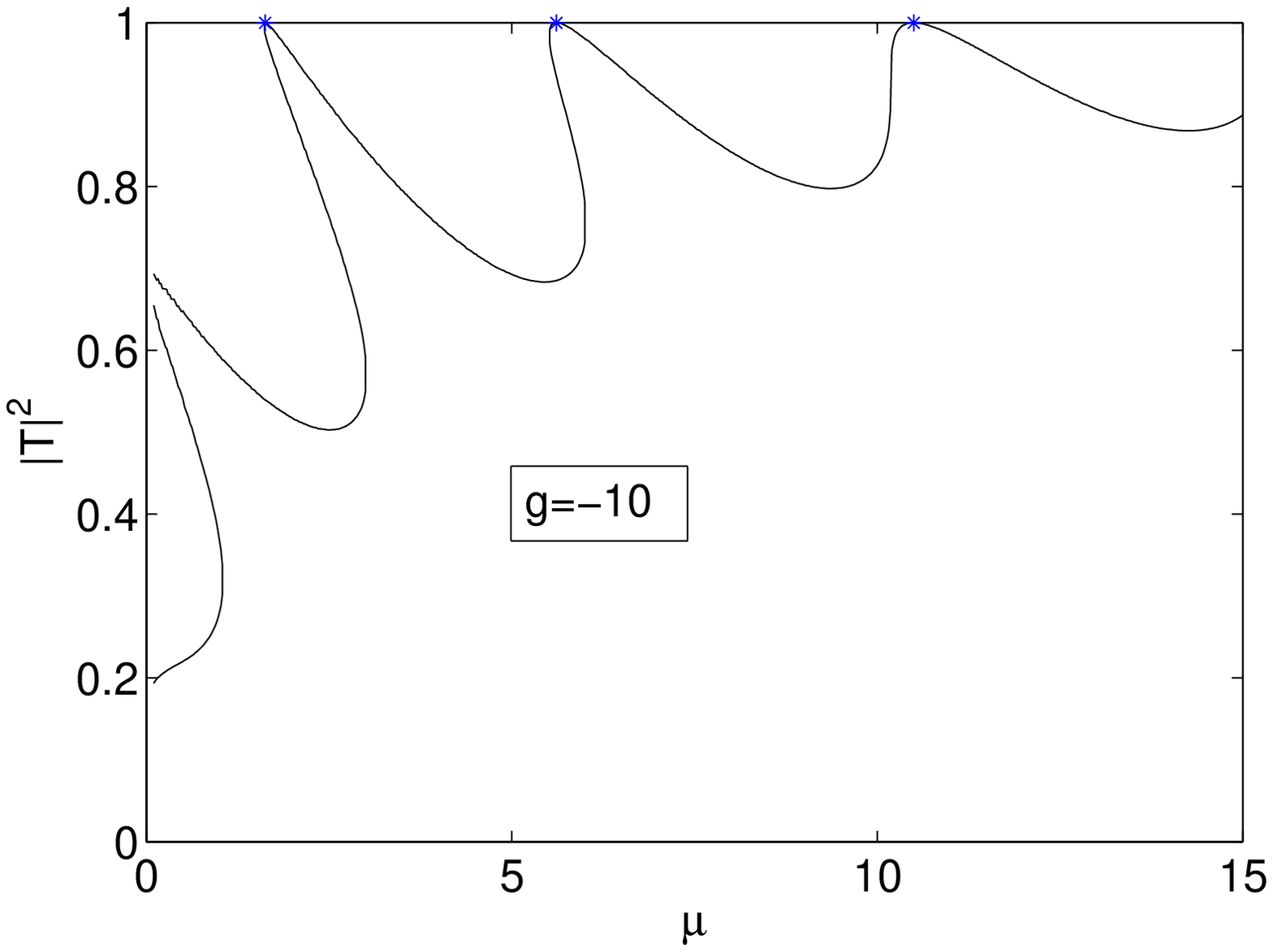}

\caption{\label{fig-T_g-2} {(Color online) Transmission coefficients for different
attractive nonlinearities for the square well potential
$V_0=-10$ , $a=2$.
The positions of the resonances, calculated with the formula
(\ref{muR}), are marked by \textcolor{blau}{'*'}.
}}
\end{figure}

\begin{figure}[htb]
\centering
\includegraphics[width=8cm,  angle=0]{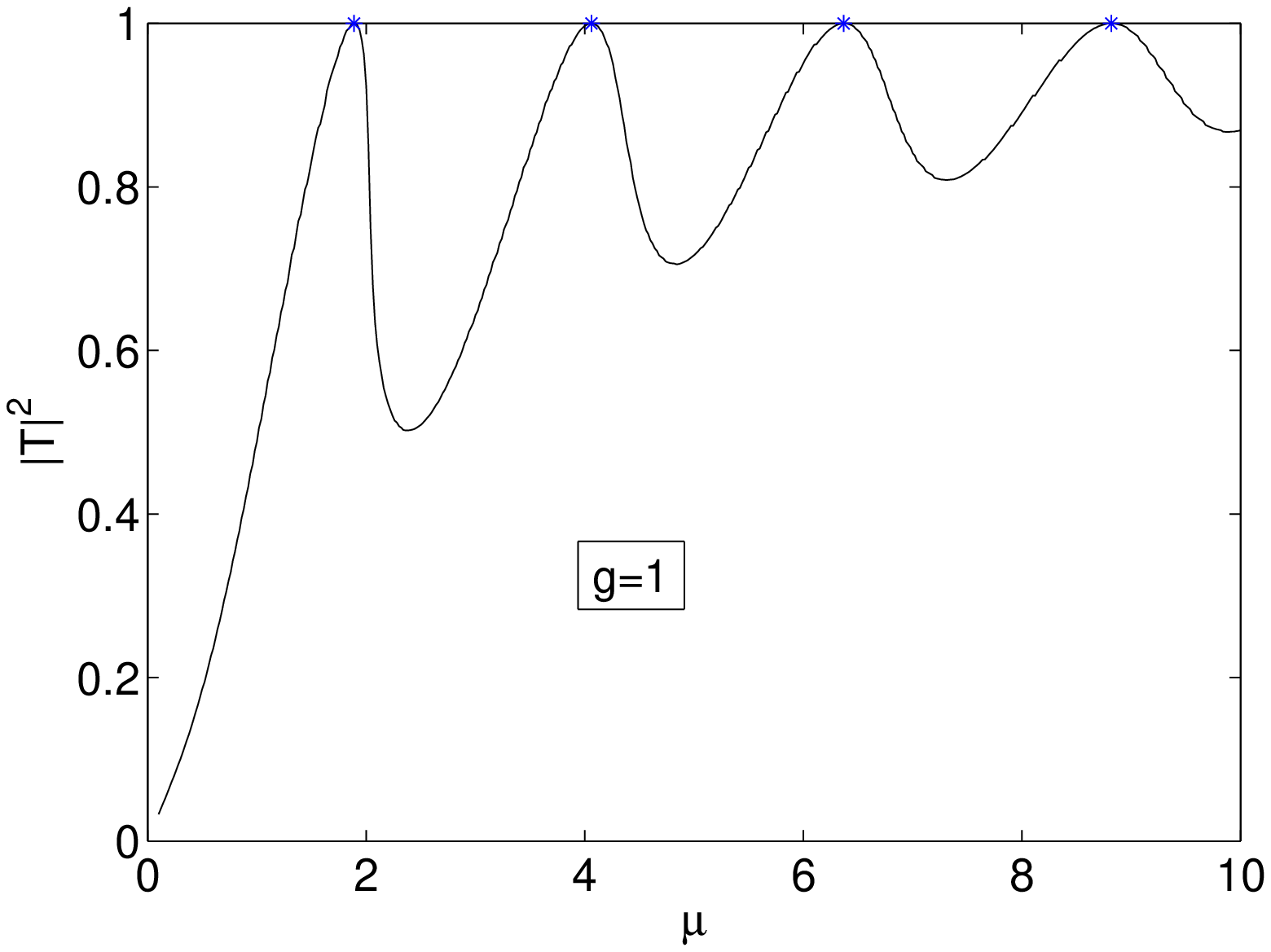}
\includegraphics[width=8cm,  angle=0]{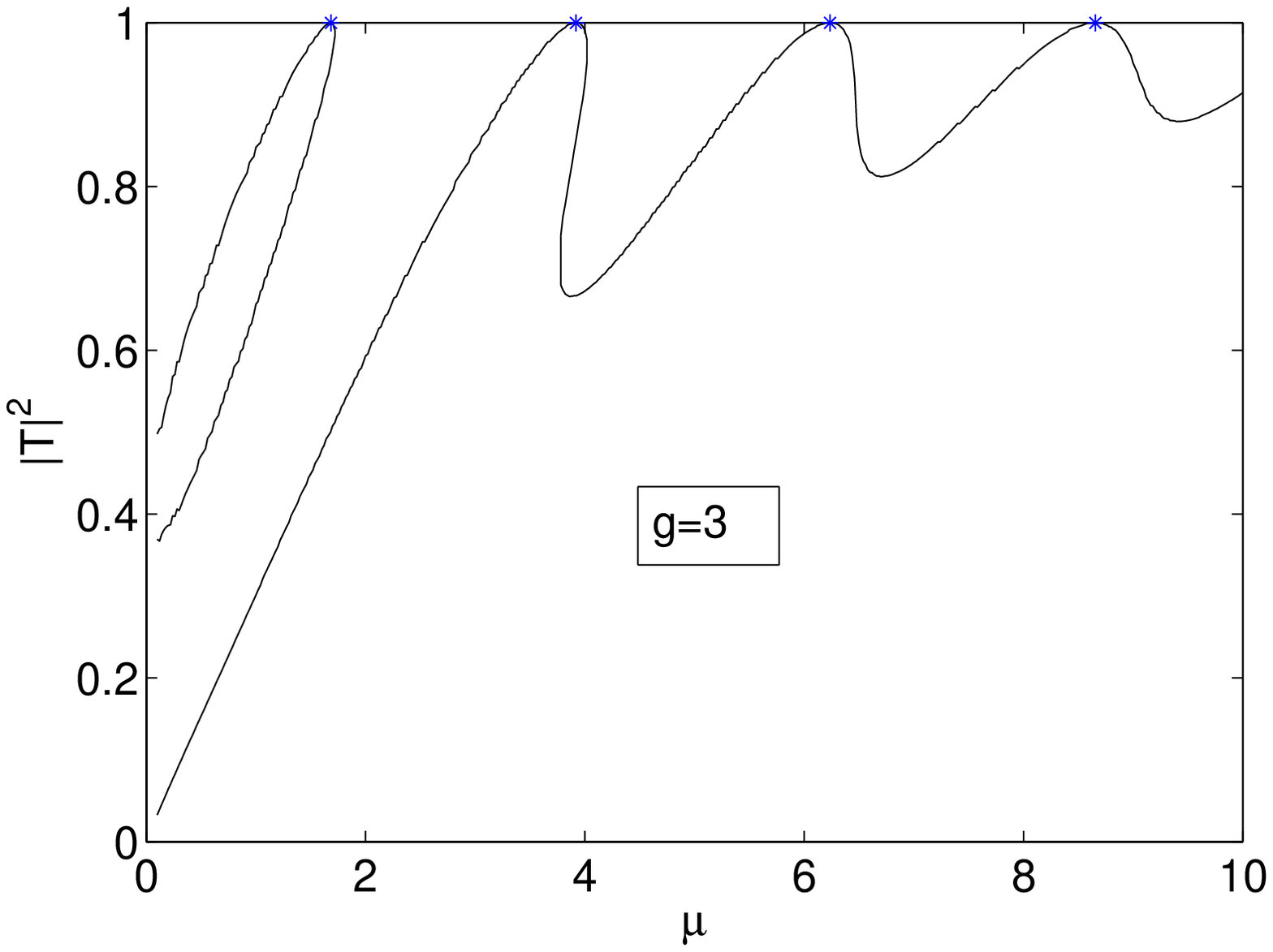}


\caption{\label{fig-T_g+} {(Color online) Transmission coefficients for different
repulsive nonlinearities for the square well potential
$V_0=-12$ , $a=4$.
The positions of the resonances, calculated with the formula
(\ref{muR}), are marked by \textcolor{blau}{'*'}.
}}
\end{figure}

\begin{figure}[htb]
\centering

\includegraphics[width=8cm,  angle=0]{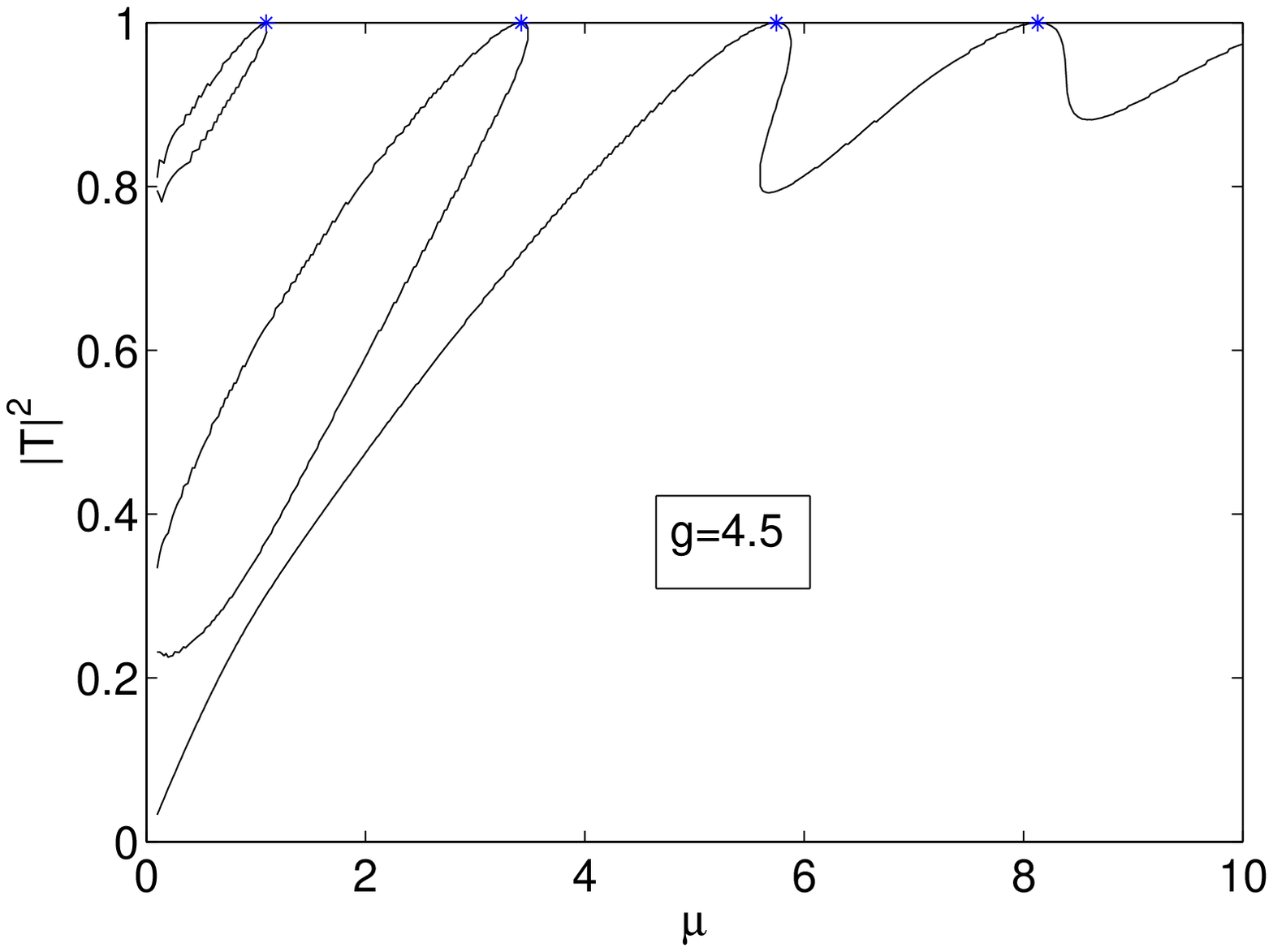}
\hspace{5mm}
\includegraphics[width=8cm,  angle=0]{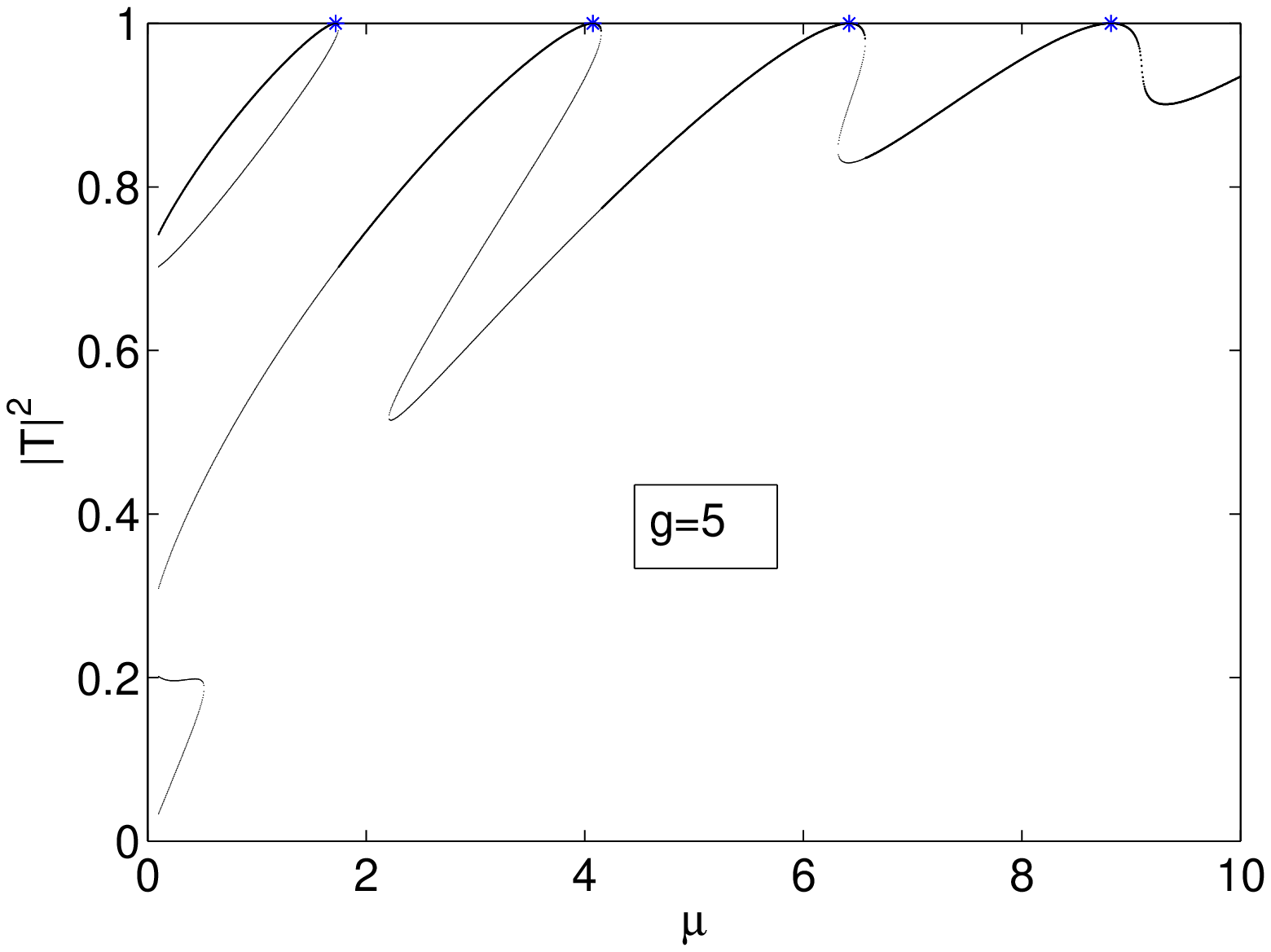}

\caption{\label{fig-T_g+2} {(Color online) Transmission coefficients for different
repulsive nonlinearities for the square well potential
$V_0=-12$ , $a=4$.
The positions of the resonances, calculated with the formula
(\ref{muR}), are marked by \textcolor{blau}{'*'}.
}}
\end{figure}

\begin{figure}[htb]
\centering
\includegraphics[width=8.5cm,  angle=0]{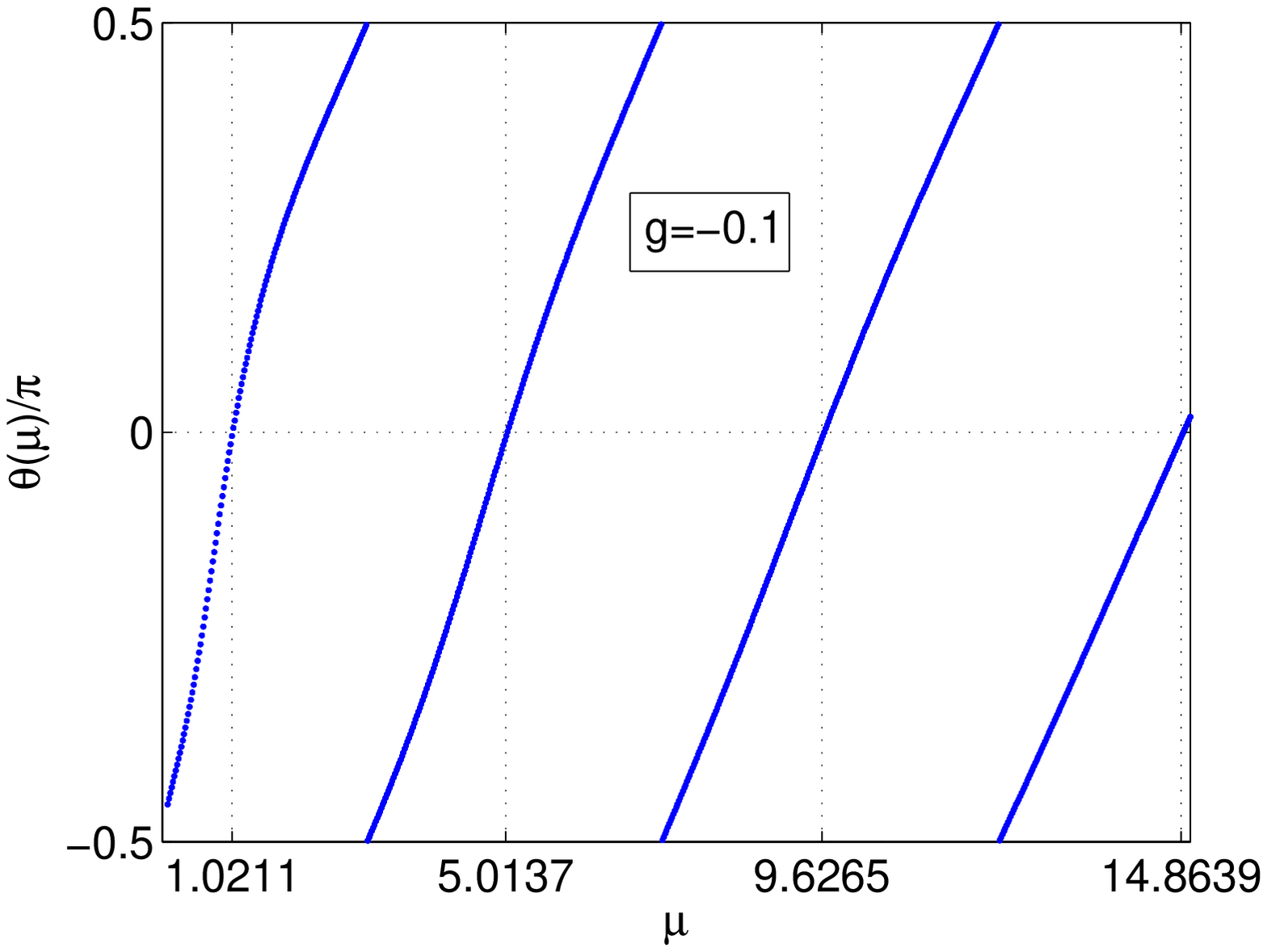}
\hspace{5mm}
\includegraphics[width=8cm,  angle=0]{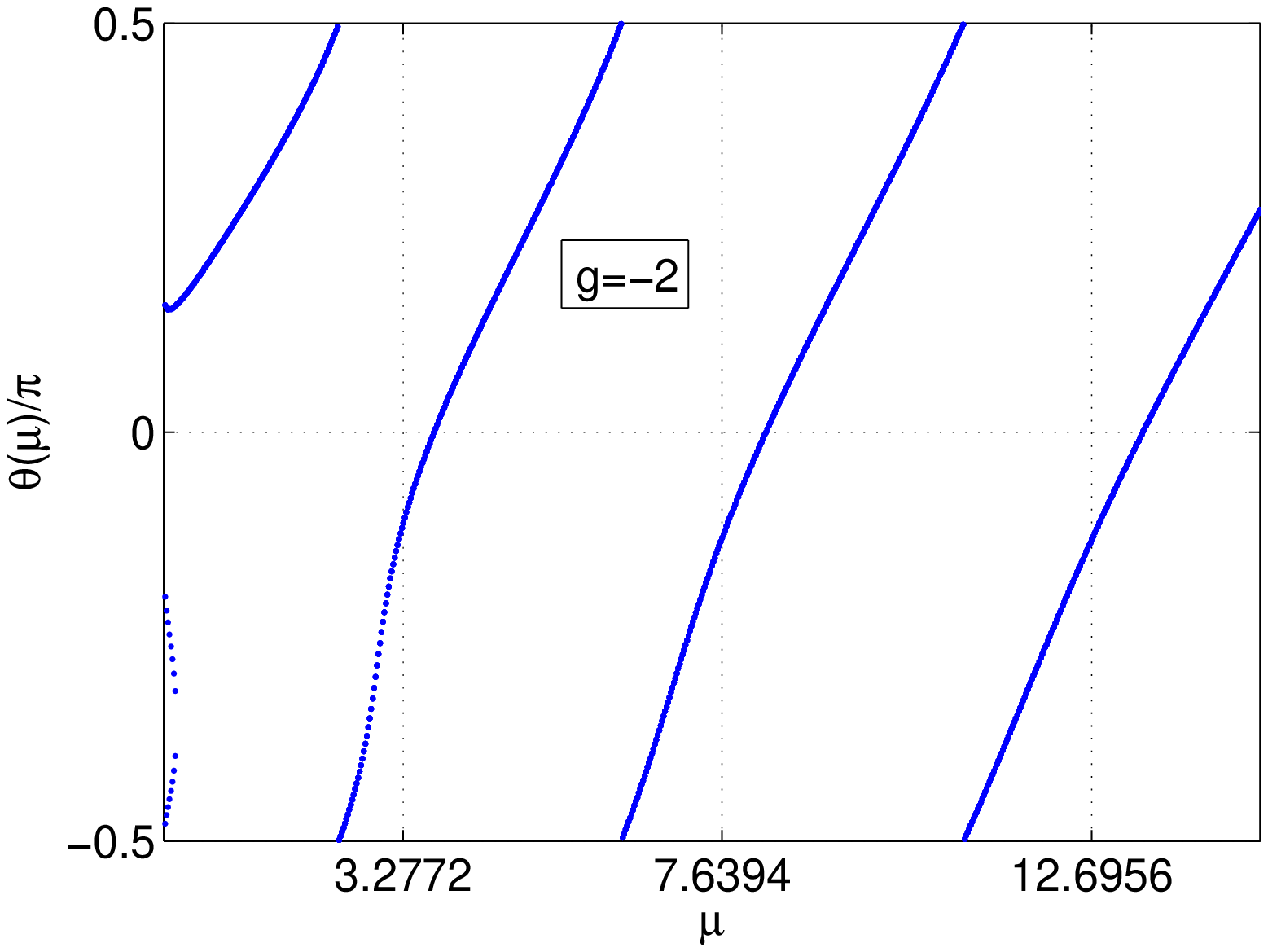}


\caption{\label{fig-Phase1} {(Color online) Scattering phases for different attractive
nonlinearities for the potential $V_0=-10$, $a=2$.
The positions of the resonances, calculated with the formula
(\ref{muR}), are marked by dotted vertical lines.
}}
\end{figure}

\begin{figure}[htb]
\centering

\includegraphics[width=8cm,  angle=0]{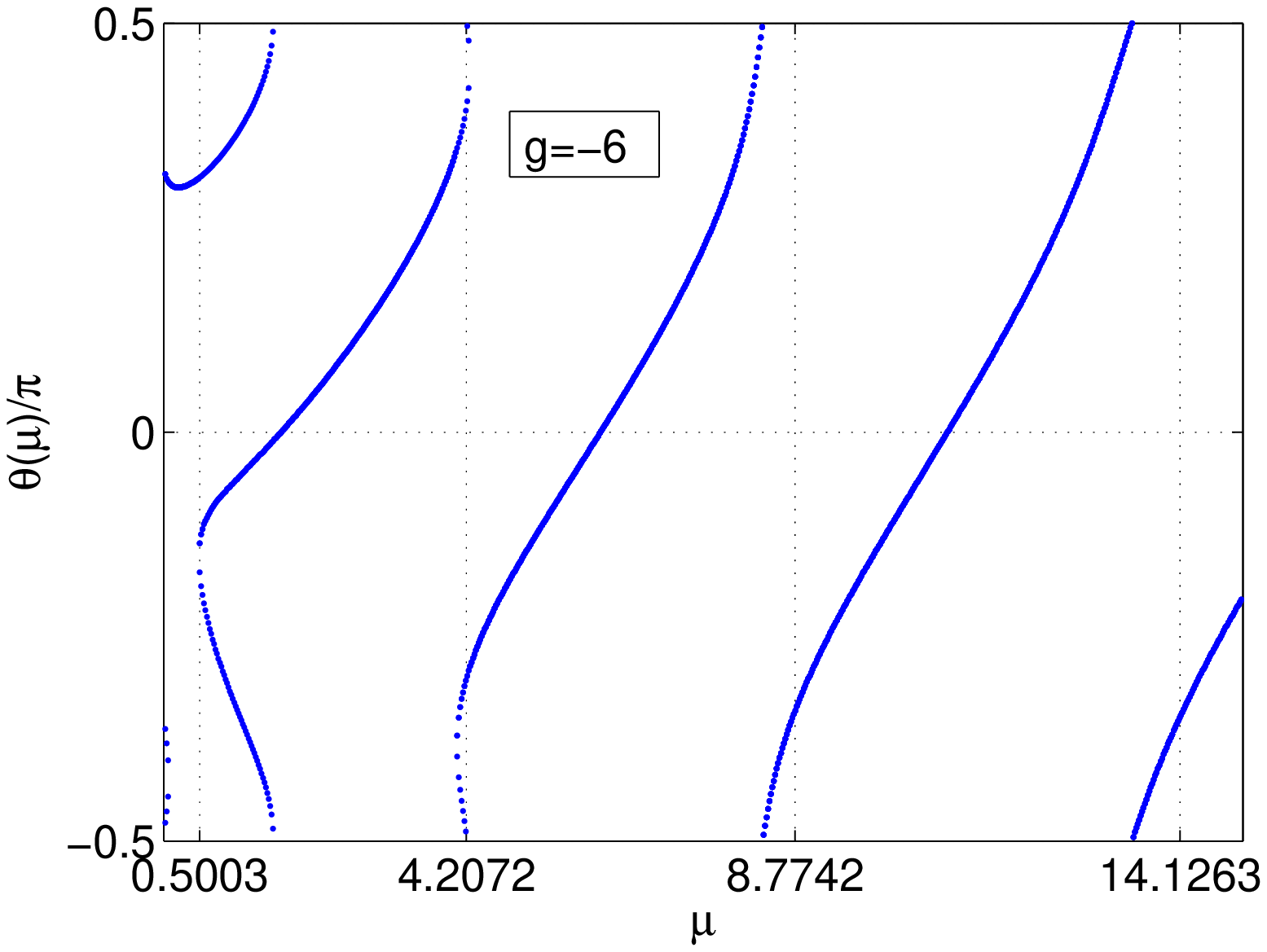}
\hspace{5mm}
\includegraphics[width=8cm,  angle=0]{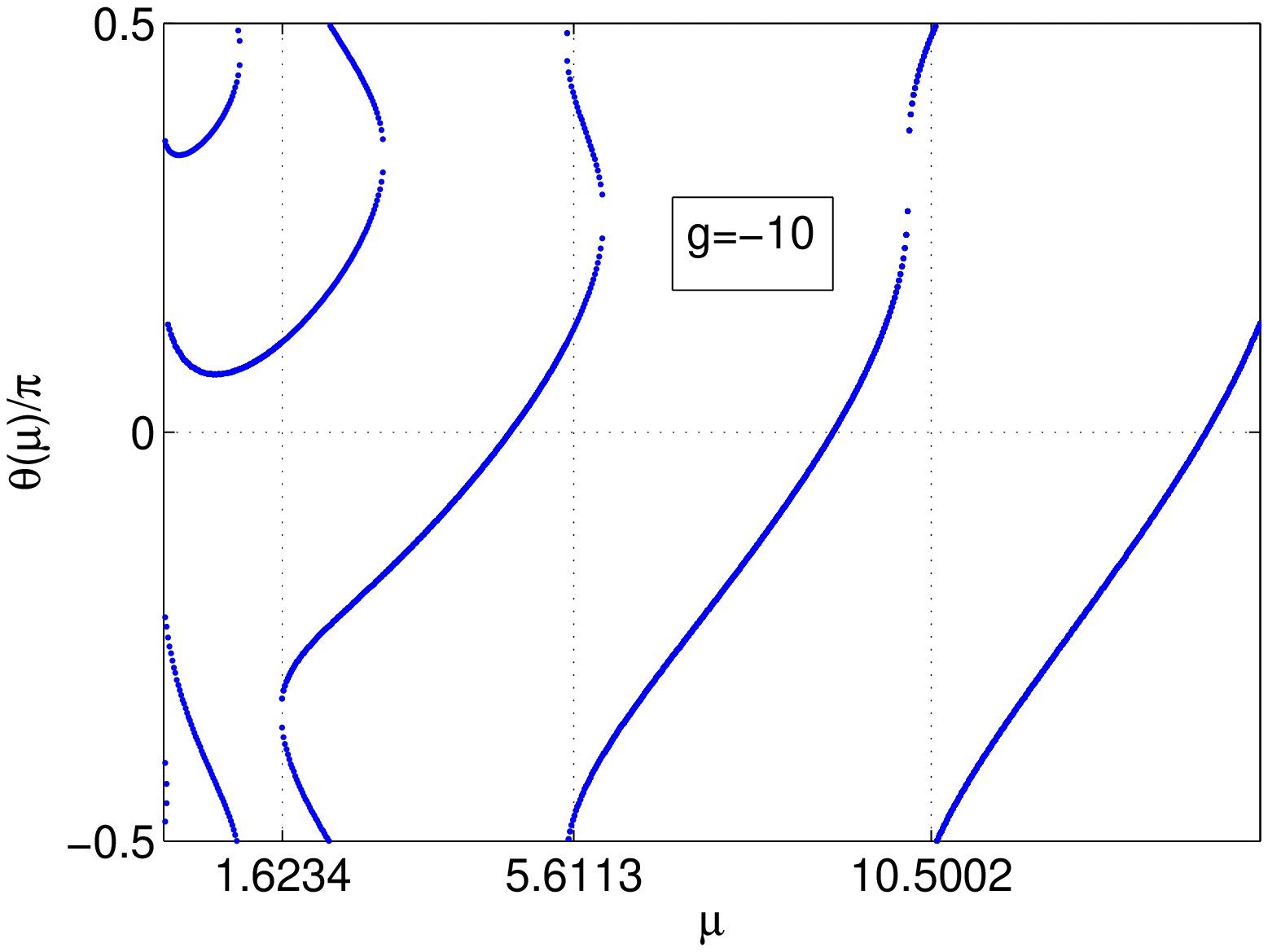}

\caption{\label{fig-Phase1_2} {(Color online) Scattering phases for different attractive
nonlinearities for the potential $V_0=-10$, $a=2$.
The positions of the resonances, calculated with the formula
(\ref{muR}), are marked by dotted vertical lines.
}}
\end{figure}

\begin{figure}[htb]
\centering
\includegraphics[width=8cm,  angle=0]{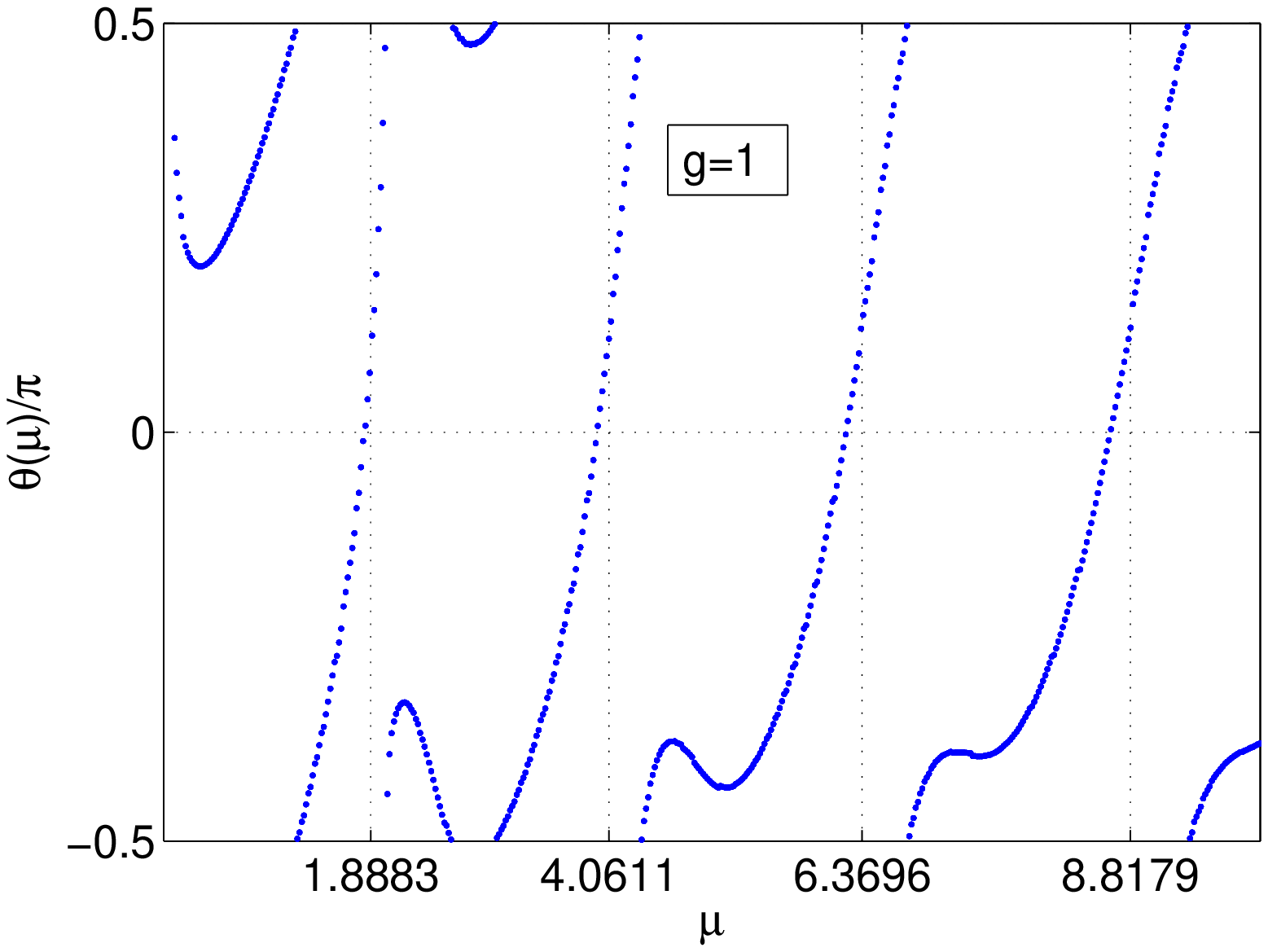}
\hspace{5mm}
\includegraphics[width=8cm,  angle=0]{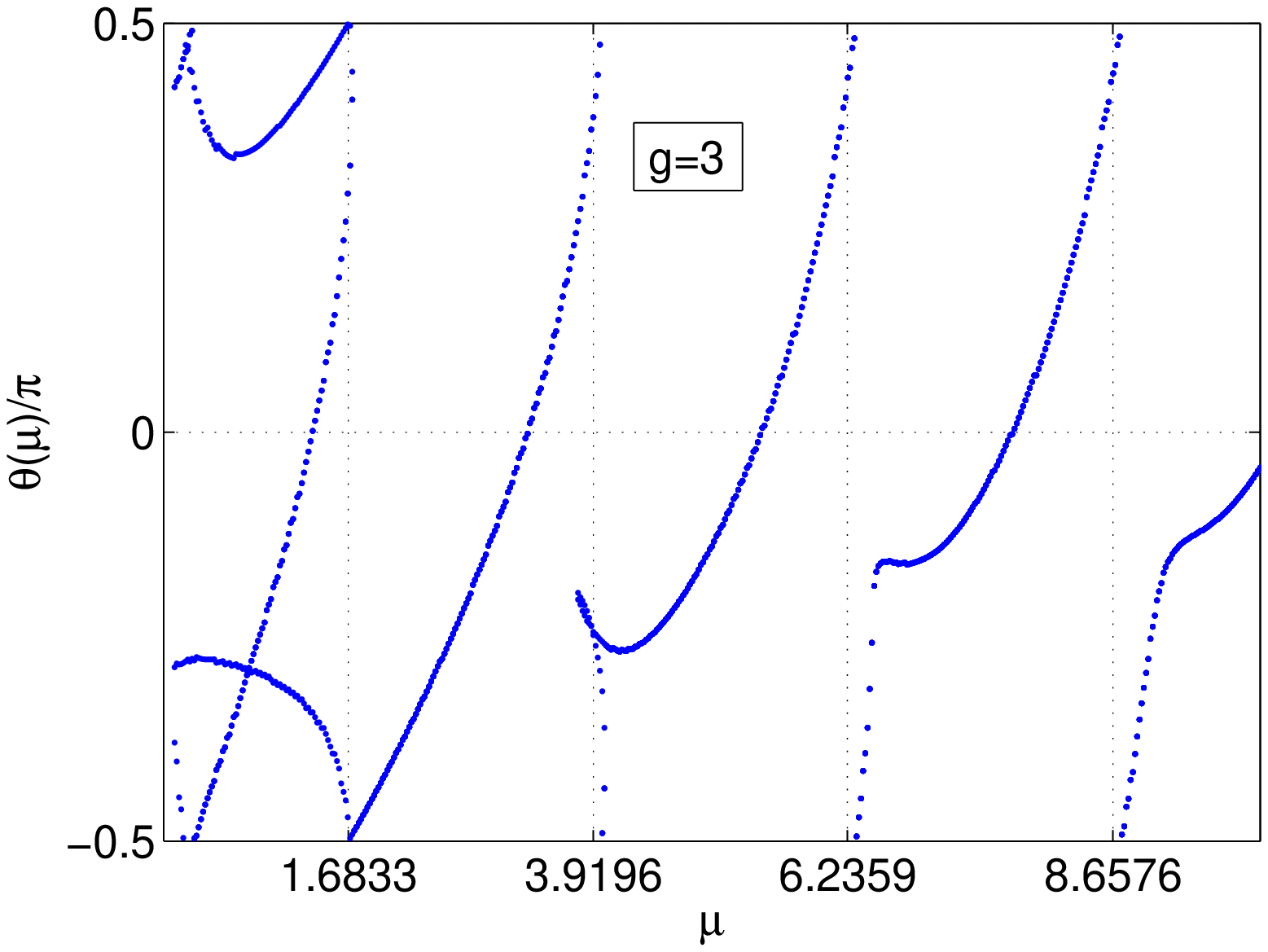}

\caption{\label{fig-Phase2} {(Color online) Scattering phases for different repulsive
nonlinearities for the potential $V_0=-12$, $a=4$.
The positions of the resonances, calculated with the formula
(\ref{muR}), are marked by dotted vertical lines.
}}
\end{figure}

\begin{figure}[htb]
\centering

\includegraphics[width=8cm,  angle=0]{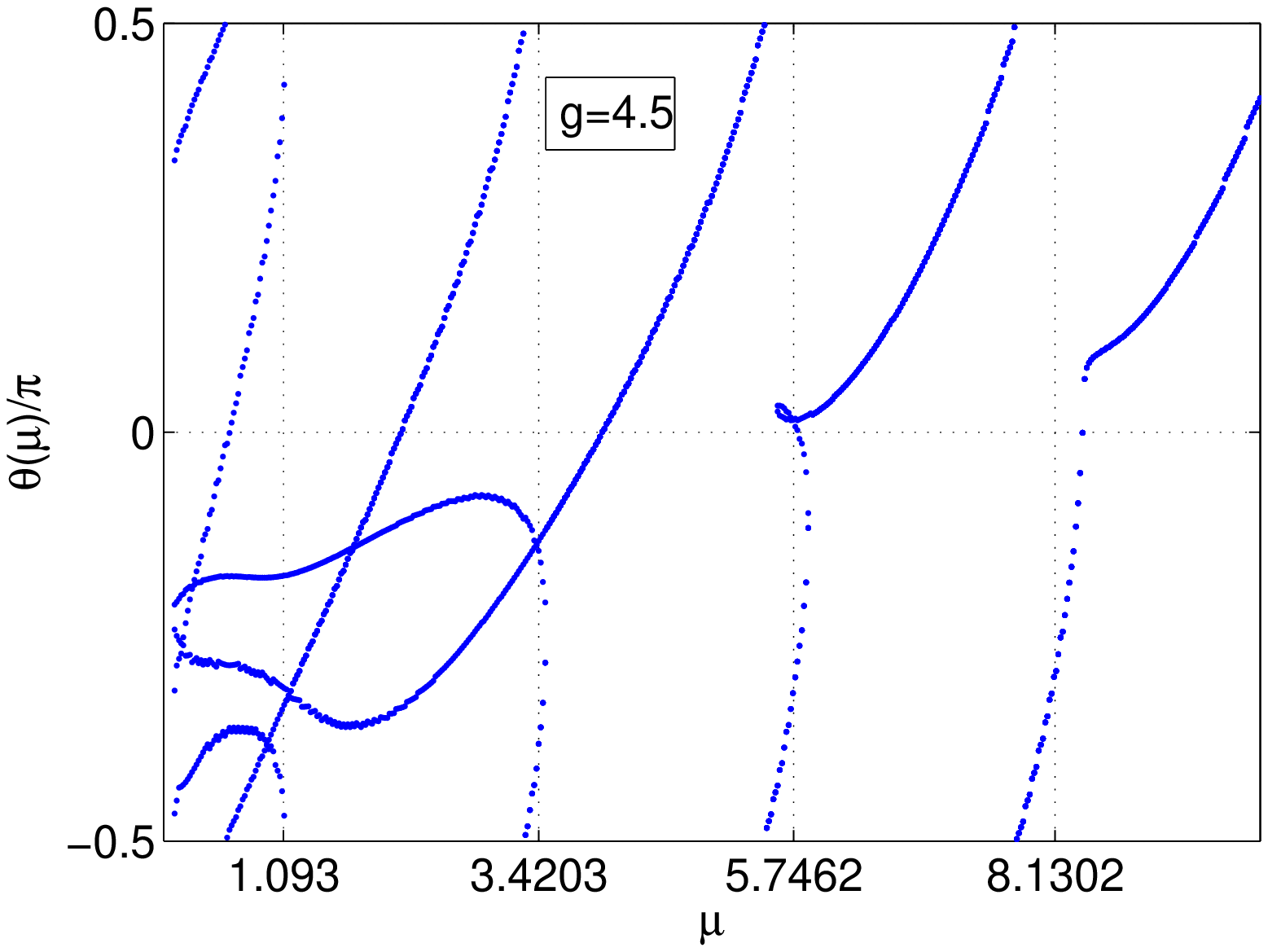}
\hspace{5mm}
\includegraphics[width=8cm,  angle=0]{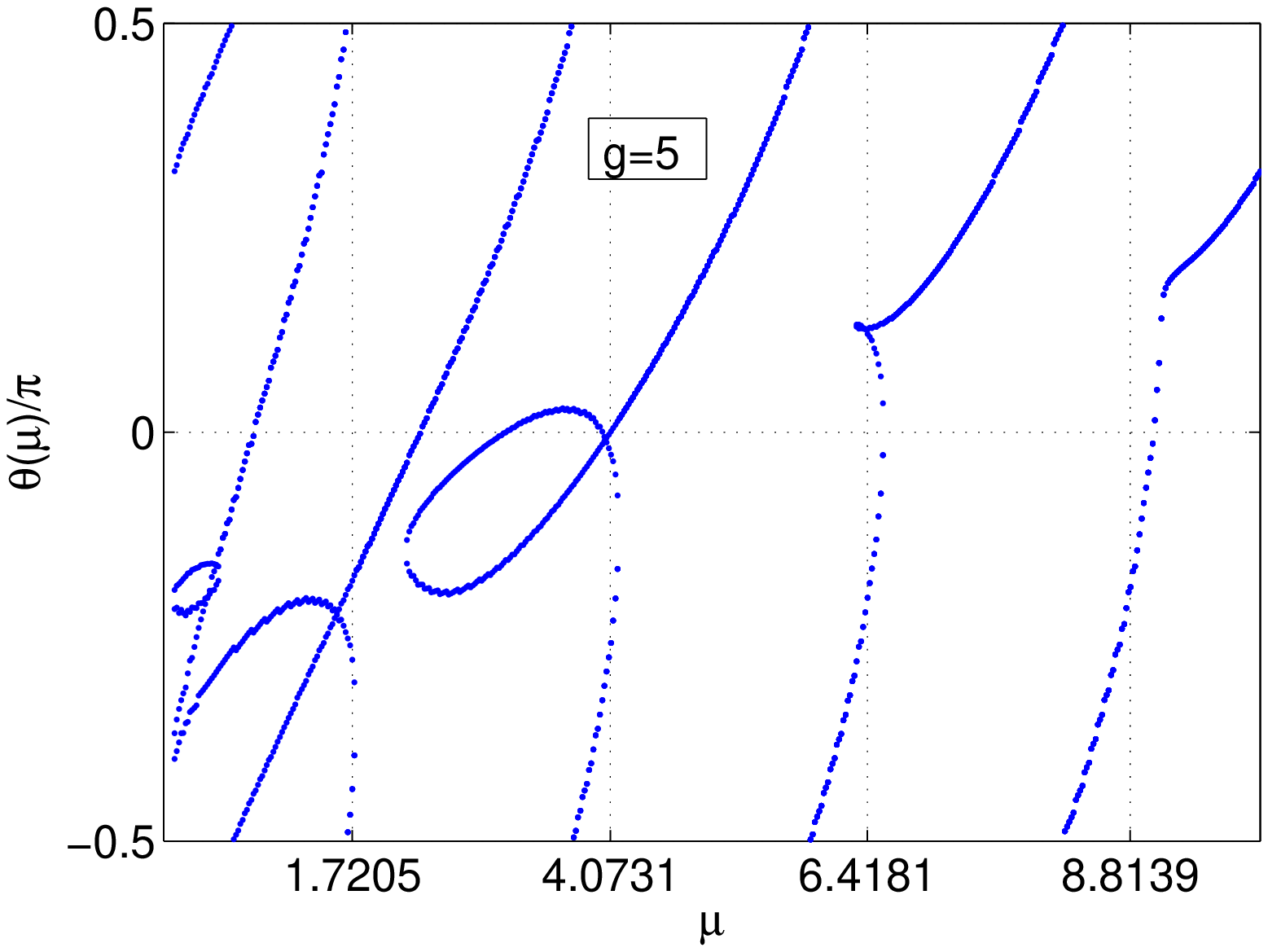}
\caption{\label{fig-Phase2_2} {(Color online) Scattering phases for different repulsive
nonlinearities for the potential $V_0=-12$, $a=4$.
The positions of the resonances, calculated with the formula
(\ref{muR}), are marked by dotted vertical lines.
}}
\end{figure}

Figures \ref{fig-WFQ-} and \ref{fig-WFQ+} show the squared modulus of resonance wavefunctions for attractive respectively repulsive nonlinearities in the region of the potential well $|x| \le a$ where scaled units with $\hbar=m=1$ are used, as for all figures and numerical calculations in this paper. The functions are symmetric with respect to $x=0$ and have no nodes. The quantum number $n$ indicates the number of the minima. The smaller the chemical potential $\mu_R(n)$ is, the smaller is the value of $|\psi(x)|^2$ at its minima.

Figures \ref{fig-T_g-} and \ref{fig-T_g-2} show the transmission coefficient $|T(\mu)|^2$ in dependence of the chemical potential $\mu$ for the potential $V_0=-10$ and $a=2$ for different attractive nonlinearities $g<0$. For small values of $|g|$, the behaviour of the transmission coefficient is very similar to the linear system (see Fig.~\ref{fig-T_lin}). With increasing attractive interaction, the curves bend more and more to the left so that the resonances are shifted towards smaller values of $\mu$. This shift is mainly determined by the term $\frac{3}{2}g|A|^2$ in Eq.~(\ref{muR}). For $g=-2$, the lowest resonance so far has disappeared into the bound state regime $\mu<0$. Between $g=-6$ and $g=-8$ the next resonance disappears. It can be shown that these resonances become bound states of the system (see section \ref{AT_bs}). For higher values of $|g|$ the bending of the curve in the vicinity of the resonances leads to a bistable behaviour. For a fixed value of $\mu$, there are now different wavefunctions which lead to different transmission coefficients $|T(\mu)|^2$. Due to the nonlinear mean field interaction the state adopted by the system  for an incoming flux from $x=-\infty$ is not only determined by the chemical potential $\mu$ and the amplitude $A$ but also by the density distribution $|\psi(x,t)|^2$ of the condensate inside the potential well so that the system has a memory (hysteresis). This will be further investigated in section \ref{dyn}.

Now we consider repulsive nonlinearities  $g>0$. Figures \ref{fig-T_g+} and \ref{fig-T_g+2} show the transmission coefficient $|T|^2$ in dependence of $\mu$ for a potential $V_0=-12$, $a=4$ for different repulsive nonlinearities. Here we observe a shift of the resonance positions towards higher values of $\mu$, since the term $\frac{3}{2}g|A|^2$ is now positive. New resonances appear for increasing $|g|$ which originate from the bound states of the linear system (see section \ref{LSE_bs}). Now the curves bend to the right and, as in the attractive case, the transmission coefficient shows a bistable behaviour. Between $g=4.5$ and $g=5$, we observe a bifurcation phenomenon at $\mu \approx 1 $ which resembles an avoided crossing. This turns out to be a beak-to-beak scenario which also occurs in two-level systems. These phenomena will be discussed in detail in a subsequent paper \cite{Grae05}.
In all transmission curves we observe that, for a fixed value of $g|A|^2$, the effect of the nonlinearity on the shape of the resonances is stronger for small values of $\mu$ than for large ones, because the kinetic energy is higher and the mean-field energy has a comparatively smaller influence. This also manifests itself in formula (\ref{muR}) for $\mu_R$ where the term proportional to $n^2$ counteracts the dominant term $\frac{3}{2}g|A|^2$.

Figures \ref{fig-Phase1}-\ref{fig-Phase2_2} show scattering phases in dependence of the chemical potential $\mu$ which have been calculated by numerical integration of Eq.~(\ref{SPhase}) for the parameters of the figures \ref{fig-T_g-}-\ref{fig-T_g+2}. For weak nonlinearities the scattering phase behaves similar to the scattering phase of the linear system which is zero at the positions $\mu=\mu_R$ of the resonances. For stronger nonlinearities the resonance scattering phase increasingly deviates from zero.
Since the squared modulus of the wavefunction enters Eq.~(\ref{SPhase}), the scattering phase $\theta(\mu)$ inherits the bistable behaviour of the transmission coefficient $|T(\mu)|^2$. Wherever beak-to-beak structures occur in the transmission coefficient the scattering phase shows loops.

\section{Linear case: bound states}
\label{LSE_bs}
Now we consider bound states of the square well potential. Again we first have a brief look at the linear system. Bound states are found for energies $V_0<E<0$. The area $|x|>a$ outside the potential well is classically forbidden and the wavefunction decays exponentially with the coefficient
\be
   \kappa=\sqrt{-2mE}/\hbar .
\ee
Inside the well it still oscillates with the wavenumber
\be
   q=\sqrt{2m(E-V_0)}/\hbar .
\ee
Due to the symmetry $V(x)=V(-x)$ of the potential we can choose eigenfunctions which are either of odd or even parity.
For even parity we make the ansatz
\be
 \psi_+(x)=  \left\{ \begin{array}{cc}
                            \gamma_+ \, \re^{\kappa x}   &    x<-a \\
                       \eta_+ \, \cos(qx)         &  |x| \le a\\
                            \gamma_+ \, \re^{-\kappa x}  &     x>a \\
                    \end{array}
              \right.
\ee
and for odd parity
\be
 \psi_-(x)=  \left\{ \begin{array}{cc}
                         \gamma_- \, \re^{\kappa x}   &    x<-a \\
                       \eta_- \, \sin(qx)         &  |x| \le a\\
                           -\gamma_- \, \re^{-\kappa x}  &     x>a \\
                    \end{array}
              \right. \, .
\ee
As in the case of scattering states the wavefunction and its derivative must be continuous at $x=\pm a$.
For even parity we get
the condition
\be
       q \tan(qa)=\kappa,
       \label{tanqa}
   \ee
which determines the energy of the bound states.
For odd parity we obtain
   \be
       q \cot(qa)=-\kappa
       \label{cotqa} \, .
   \ee

The transcendental equations (\ref{tanqa}) and (\ref{cotqa}) can be solved numerically. One can show \cite{Nolt92} that the number of symmetric or antisymmetric solutions is given by
\be
   N_+=\left[ \frac{R_0}{\pi}\right] _>
\quad \text{and} \quad
   N_-=\left[ \frac{R_0}{\pi}-\frac{1}{2}\right ]_>
   \label{Nminus} \, ,
\ee
respectively, where $[y]_>$ denotes the smallest integer greater than $y$ and $R_0$ is given by
\be
   R_0=\sqrt{\frac{2m}{\hbar^2}a^2V_0} \, .
\ee

\section{Attractive nonlinearity: bound states}
\label{AT_bs}

In this section, we consider bound states of the nonlinear system described by the equations (\ref{sq_innen}) and (\ref{sq_aussen}) for attractive nonlinearities $g<0$. In analogy to the linear case, bound states occur only for chemical potentials $\mu<0$. As in the linear case, we look for solutions of even and odd parity. We make the ansatz
\be
 \psi(x)_+=  \left\{ \begin{array}{cc}
                           \gamma_+ \, \re^{\kappa x}   &    x<-a \\
                 I_+ \, \mbox{cn}(\varrho_+ x |p_+)     &    |x| \le a\\
                           \gamma_+ \, \re^{-\kappa x}  &     x>a \\
                    \end{array}
              \right.
\ee
for even and
\be
 \psi(x)_-=  \left\{ \begin{array}{cc}
                                  \gamma_- \, \re^{\kappa x}   &    x<-a \\
                 I_- \mbox{cn}(\varrho_- x +K(p_-)|p_-)     &  |x| \le a\\
                                 -\gamma_- \re^{-\kappa x}  &     x>a \\
                    \end{array}
              \right.
\ee
for odd parity with
\be
  \kappa=\sqrt{-2m\mu}/\hbar .
  \label{kappa_nl}
\ee

The solutions inside the potential well $|x| \le a$ are given by Jacobi elliptic functions $\cn$ with amplitudes
\be
    I_\pm=\sqrt{-\frac{\hbar^2 \varrho_\pm^2 p_\pm}{g m}}
    \label{I_pm}
\ee
and wavenumbers
\be
   \varrho_\pm =\sqrt{\frac{2m(\mu-V)}{\hbar^2(1-2p_\pm)}}.
   \label{rho_pm}
\ee

\begin{enumerate}
\item{symmetric solutions} \\
The continuity of the wavefunction and its derivative at $x=\pm a$ yields with the abbreviation $u=\varrho_+ a)$
 \be
      \gamma_+ \, \re^{-\kappa a}=I_+ \, \mbox{cn}(u|p_+)
      \label{gamma_p_I_p}
   \ee
   \be
      \kappa \gamma_+ \, \re^{-\kappa a}= I_+ \, \varrho_+ \, \mbox{sn}( u|p_+)\mbox{dn}(u|p_+).
      \label{5.135}
   \ee
By inserting Eq.~(\ref{gamma_p_I_p}) into Eq.~(\ref{5.135}) we get the condition
 \be
      \kappa \,  \mbox{cn}(u|p_+) -\varrho_+ \, \mbox{sn}(u|p_+) \mbox{dn}(u|p_+) = 0
     \label{tan_nl}
   \ee
for symmetric bound states. In the limit $p \rightarrow 0$, the Jacobi elliptic functions converge to trigonometric functions and we regain the condition (\ref{tanqa}) of the linear system.

\item{antisymmetric solutions} \\
In analogy to the case of even parity we obtain the condition
  \be
      \kappa \, \mbox{cn}(u|p_-) -\varrho_- \, \mbox{sn}(u|p_-) \mbox{dn}(u|p_-) = 0
      \label{cot_nl}
   \ee
for antisymmetric bound states with $u=\varrho_- \, a+K(p_-)$, which also converges to Eq.~(\ref{cotqa}) in the limit $p  \rightarrow 0$.
\end{enumerate}

Now we show explicitly that the resonances discussed in section \ref{NLSE_Scat} become bound states for a sufficiently strong attractive mean-field interaction. Due to the behaviour of the wavefunction outside the potential well, transitions from resonances to bound states occur at $\mu=0$. For $\mu=0$ the condition (\ref{tan_nl}) for symmetric bound states becomes
\be
   \mbox{sn}(\varrho_+ \, a|p_+)=0
\ee
which is solved by
\be
   \varrho_+ =2j_+ \,K(p_+)/a
   \label{rho_p_0}
\ee
with integer $j_+$. Inserting Eq.~(\ref{rho_p_0}) into (\ref{rho_pm}) yields
\be
   (1-2p_+)K^2(p_+)=\frac{m|V_0|a^2}{2\hbar^2j_+^2} \ .
   \label{p_plus_ex}
\ee
For small values of $p_+$ we can make a Taylor approximation and obtain
\be
    p_+ = -\frac{24}{23}+\sqrt{ \left(\frac{24}{23}\right)^2 + \zeta_+} + \mathcal{O}(p_+^3)
    \label{p_plus_ap}
\ee
as an approximate solution for the parameter $p_+$ where the transition occurs. Here we have introduced the abbreviation
\be
   \zeta_+=\frac{32}{23}\left(1-\frac{2m |V_0| a^2}{\hbar^2 \pi^2 j_+^2}\right)=\frac{32}{23}\left(1-\frac{(R_0/\pi)^2}{j_+^2}\right) \   .
\ee
The quantity $R_0/\pi$ in $\zeta_+$ determines the number of states $N_+=\left[R_0/\pi \right]_>$ for $g=0$ (see sction \ref{LSE_bs}). Since $p_+$ must be positive, $\zeta_+$ must be positive as well and we obtain the condition
\be
   j_+ \ge N_+ \ .
\label{j_gr_Np}
\ee

Next we want to demonstrate that the wavefunction of a symmetric bound state continuously merges into a resonance wavefunction with an even quantum number $n$ at $\mu=0$ if we use an appropriate normalization. To avoid confusion, all the parameters of the resonance wavefunction now carry an additional index $R$. For $\mu=0$ we have $k=\kappa=0$ (see equations (\ref{well_lin}) and (\ref{kappa_nl})) so that the squares of the wavefunctions outside the potential well are identical if we use the normalization
\be
   |\gamma_+|=|A| .
\ee
The equations (\ref{I_pm}), (\ref{gamma_p_I_p}) and (\ref{rho_p_0}) then imply
\be
   |A|^2=|I_+|^2=-\frac{4 \hbar^2 j_+^2K(p_+)^2 p_+}{g m a^2} \ .
   \label{Aq=Iq}
\ee
Inside the potential well the squared modulus of the resonance wavefunction is given by
\be
   |\psi_R(x)|^2=|A|^2 -\frac{\hbar^2K^2(p_R)n^2}{gma^2} \left[\mbox{dn}^2\left(\frac{nK(p_R)}{a}x \big|p_R\right)-1\right]
\ee
(see Eq.~(\ref{S_R})) as $g<0$ corresponds to case 1 in section \ref{NLSE_Scat}.
Because of $k=0$, the phase of the resonance wavefunction is constant and the phases of $\psi_R$ and $\psi_+$ can be chosen identical. Using Eq.~(\ref{Aq=Iq}) and the relation between the squares of the Jacobi elliptic functions \cite{Abra72,Lawd89,Bowm61} we arrive at
\begin{eqnarray}
   |\psi_R(x)|^2=&-&\frac{\hbar^2 4j_+^2K(p_+)^2 p_+}{g m a^2} -\frac{\hbar^2K^2(p_R)n^2}{gma^2}p_R  \nonumber \\
                 &\times& \left[\mbox{cn}^2\left(\frac{nK(p_R)}{a}x \big|p_R\right)-1\right] \ .
\end{eqnarray}
By comparing $|\psi_R(x)|^2$ with the squared modulus of the bound state wavefunction,
\be
   |\psi_+(x)|^2 =-\frac{\hbar^2 4j_+^2K(p_+)^2 p_+}{g m a^2}\mbox{cn}^2\left(\frac{2jK(p_+)}{a}x \big|p_+\right) \, ,
\ee
we see that $|\psi_R(x)|^2$ is equal to $|\psi_+(x)|^2$ if
\be
   p_R=p_+=:p_{T+}
   \label{p=p}
\ee
and
\be
   n=2j_+
   \label{n=2j} \, ,
\ee
where the index $T$ denotes transition.
Eq.~(\ref{Aq=Iq}) implies that the transition occurs for the nonlinear interaction strength
\be
   g_{T+}=-\frac{\hbar^2n^2K(p_{T+})^2p_{T+}}{ma^2|A|^2} .
   \label{gT}
\ee
Inserting Eq.~(\ref{n=2j}), (\ref{gT}) and (\ref{p_plus_ex}) into Eq.~(\ref{muR}) for $\mu_R$ and Eq.~(\ref{pK4_1}), which connects $g$ and $p_R$, we see that
\be
   \mu_R(p=p_{T+})=0
\ee
and that Eq.~(\ref{pK4_1}) holds. Thus we have proved that Eq.~(\ref{p=p}) is valid so that resonances of even quantum number $n$ indeed become symmetric bound states.

\begin{table}[htbp]
\centering
\begin{tabular}{|l|r|r|r|r|l|}
\hline
        & \multicolumn{2}{c|}{symmetric $(n=6)$} & \multicolumn{2}{c|}{antisymmetric $(n=7)$} \\
             & numerically exact  & approx.     & numerically exact   & approx.      \\ \hline
$p_T$          & $0.0644$       & $0.0643$    & $0.2044$          & $0.205$       \\ \hline
$g_T$          & $-1.4772$       & $-1.4749$    & $-6.9159$         & $-6.969$      \\ \hline
$g^{\rm eff}_T$    & $-0.7325$       & $-0.7331$    & $-3.3592$         & $-3.346$     \\ \hline
\end{tabular}
\caption{ {Critical parameters for the transition of the two most stable resonances of the potential $V_0=-10$, $a=2$ to bound states. The approximations (\ref{p_plus_ap}) for $p_+$ and (\ref{p_minus_ap}) for $p_-$ have been used.}}
\label{tab-Res-BS}
\end{table}

\begin{figure}[htb]
\centering
\includegraphics[width=8cm,  angle=0]{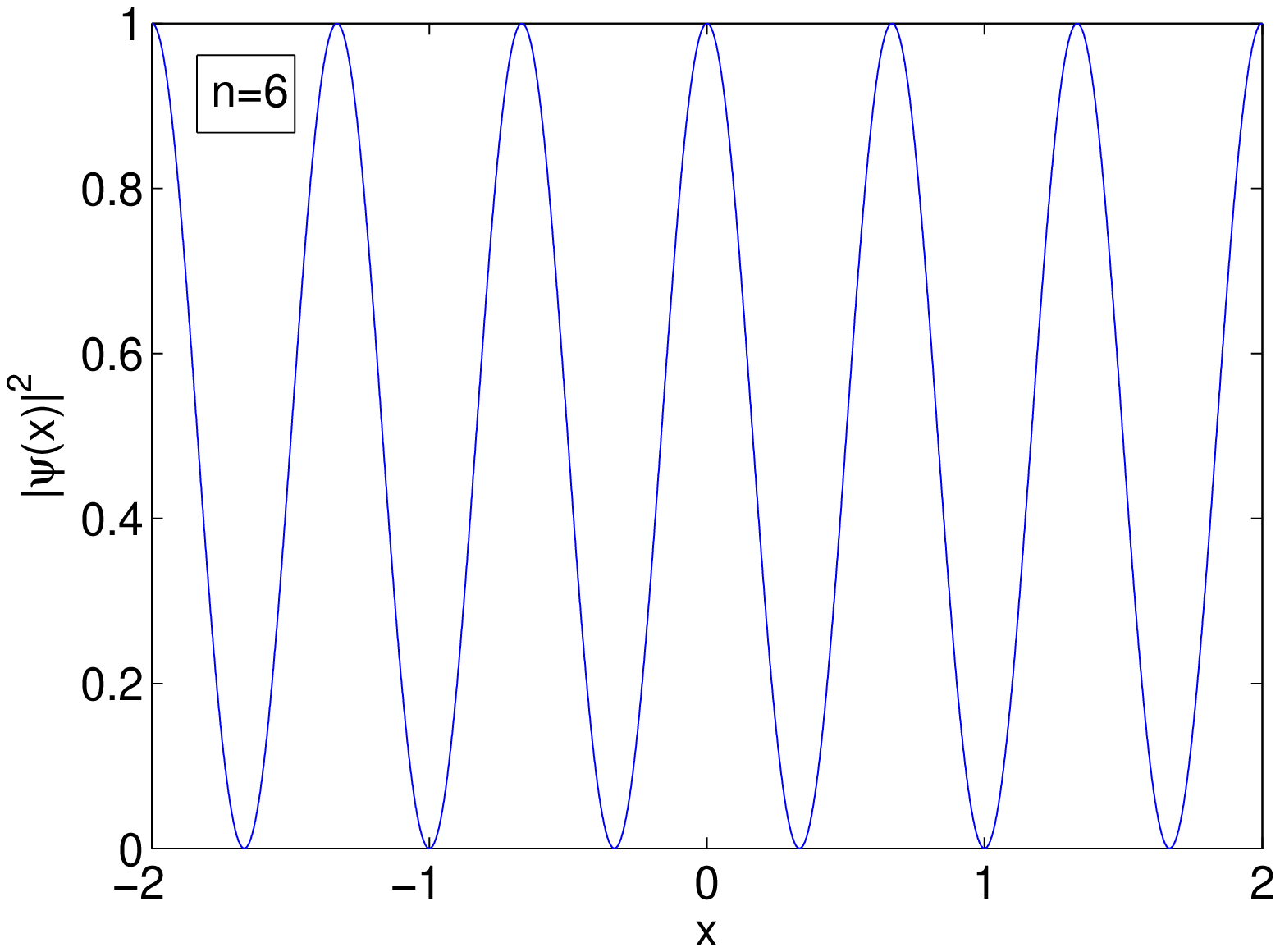}
\includegraphics[width=9cm,  angle=0]{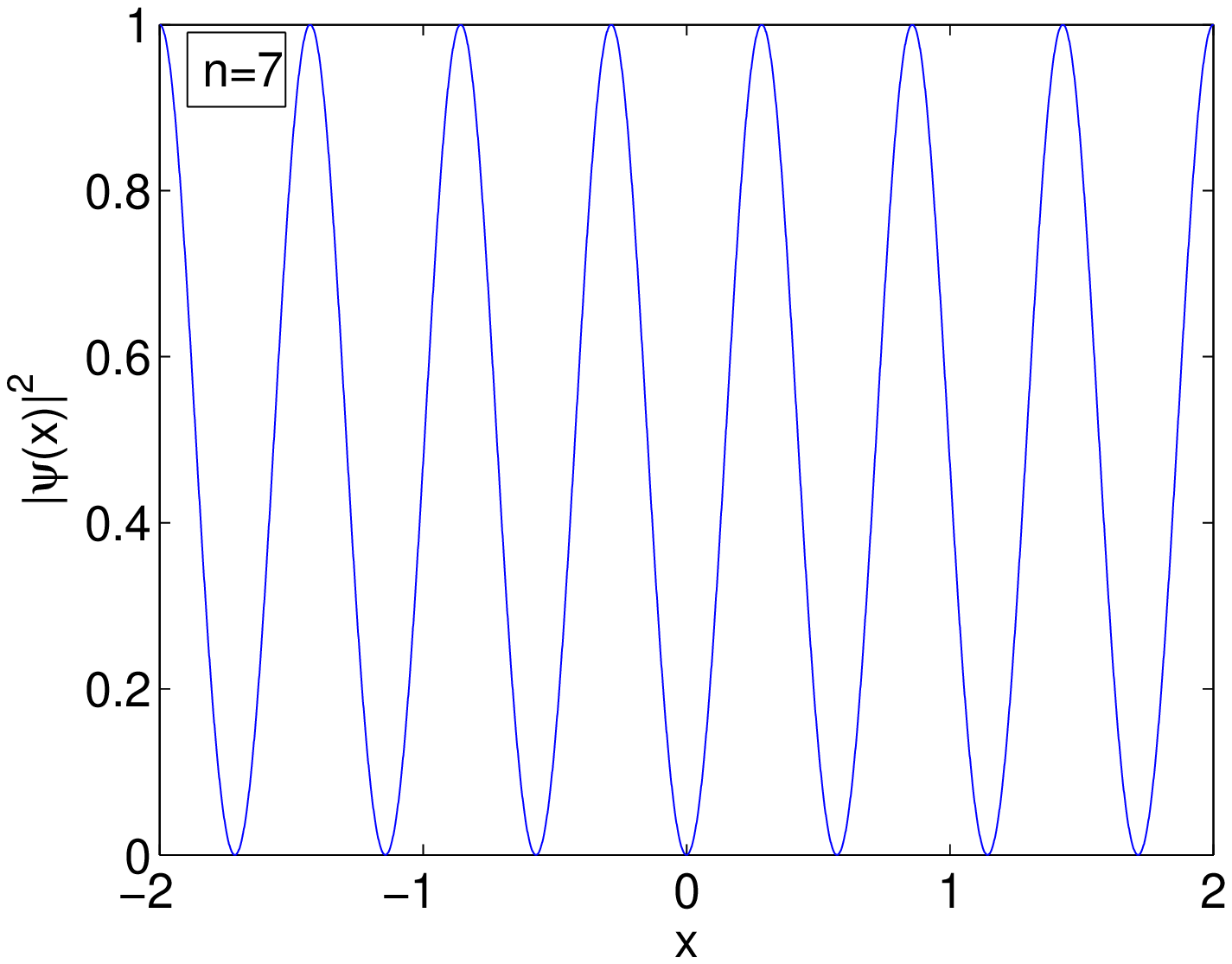}
\caption{\label{fig-WFT1} {(Color online) Squared modulus of the wavefunctions of the two most stable resonances of the potential $V_0=-10$, $a=2$ in each case exactly at the critical points for the transition from resonances to bound states (comp. table \ref{tab-Res-BS}). Here, the minima are also zeros of the wavefunction (cf. Fig.~\ref{fig-WFQ-}). The normalization $|A|=1$ has been used.}}
\end{figure}

In a similar way one can show that resonances of odd quantum number $n$ become antisymmetric bound states for a sufficiently strong attractive nonlinearity $g<0$.
We obtain (cf. Eq.~(\ref{rho_p_0}))
\be
   \varrho_- =(2j_-+1)K(p_-)/a
\ee
with integer $j_-$ and
   $n=2j_-+1$ .
If we choose the normalization
   $|\gamma_-|=|A|$,
the transition occurs at
\be
   g_{T-}=-\frac{\hbar^2n^2K(p_{T-})^2p_{T-}}{ma^2|A|^2} \, .
\ee
The parameter $p_{T-}:=p_-=p_R$ is determined by
\be
   (1-2p_{T-})K^2(p_{T-})=\frac{2m|V_0|a^2}{\hbar^2(2j_-+1)^2}
   \label{p_minus_ex}
\ee
or
\be
    p_{T-} = -\frac{24}{23}+\sqrt{ \left(\frac{24}{23}\right)^2 + \zeta_-} + \mathcal{O}(p_-^3) \, ,
    \label{p_minus_ap}
\ee
where we have introduced the abbreviation
\be
   \zeta_-=\frac{32}{23}\left(1-\frac{8m |V_0| a^2}{\hbar^2 \pi^2 (2j_-+1)^2}\right) .
\ee
The condition $p_{T-}>0$ implies $\zeta_->0$ and thus
   $j_- \ge N_- $
where  $N_-=\left[\frac{R_0}{\pi}-\frac{1}{2}\right]_>$ is the number of antisymmetric bound states for $g=0$ as defined in Eq.~(\ref{Nminus}).

\begin{figure}[htb]
\centering
\includegraphics[width=8cm,  angle=0]{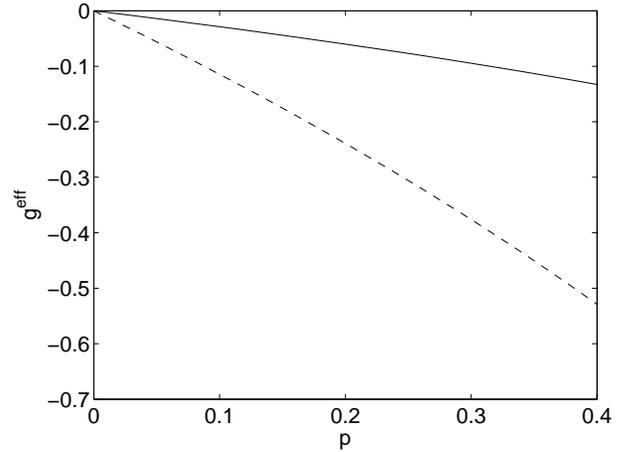}
\caption{\label{fig-geff(p)} {Dependence of the effective
nonlinear interaction strength $g^{\rm eff}_\pm$ on the
parameter $p_\pm$ for the symmetric bound state $n=0$
(solid line) and the antisymmetric bound state $n=1$ (dashed line). }}
\end{figure}

\begin{figure}[htb]
\centering
\includegraphics[width=8cm,  angle=0]{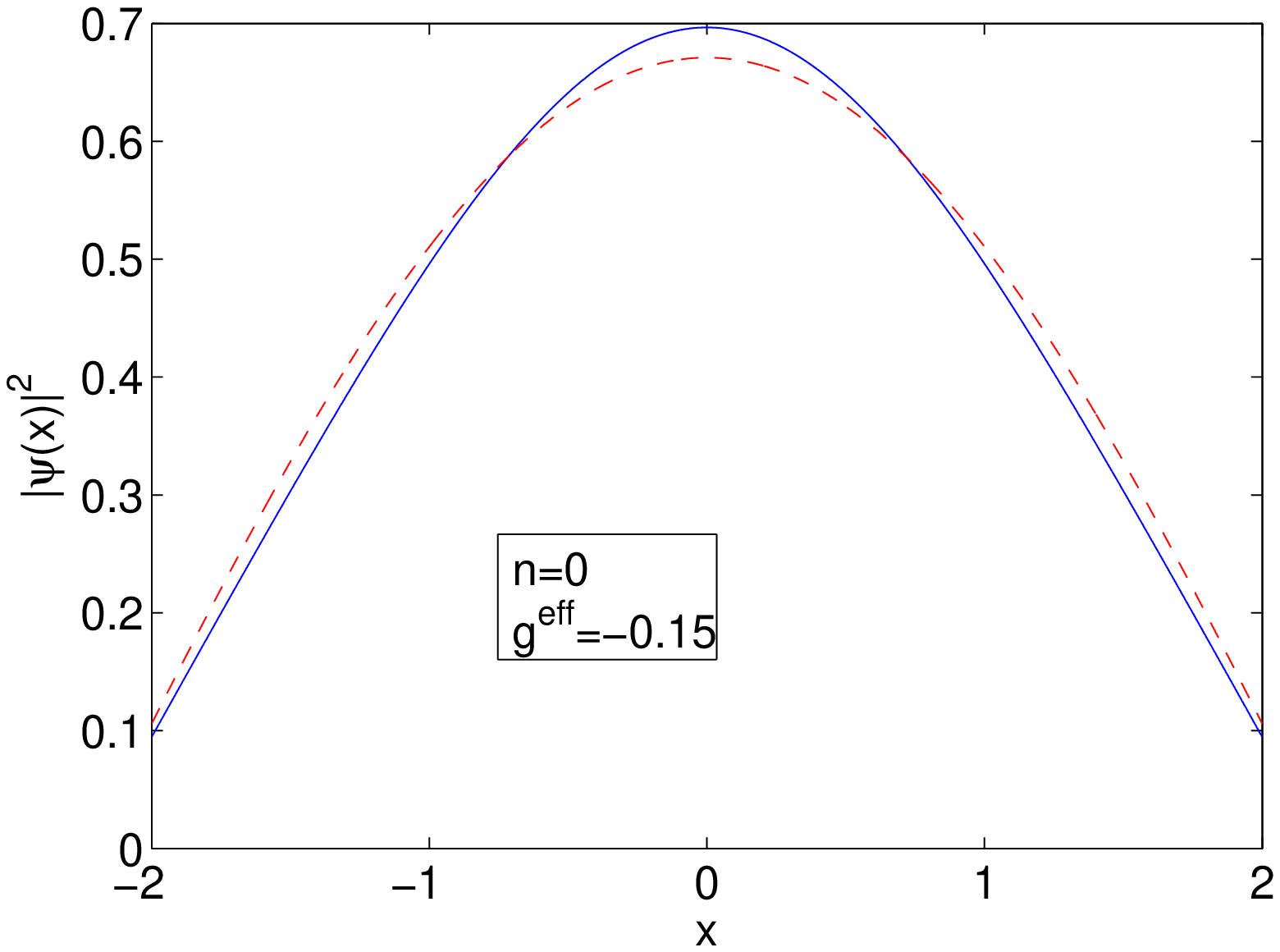}
\hspace{5mm}
\includegraphics[width=8cm,  angle=0]{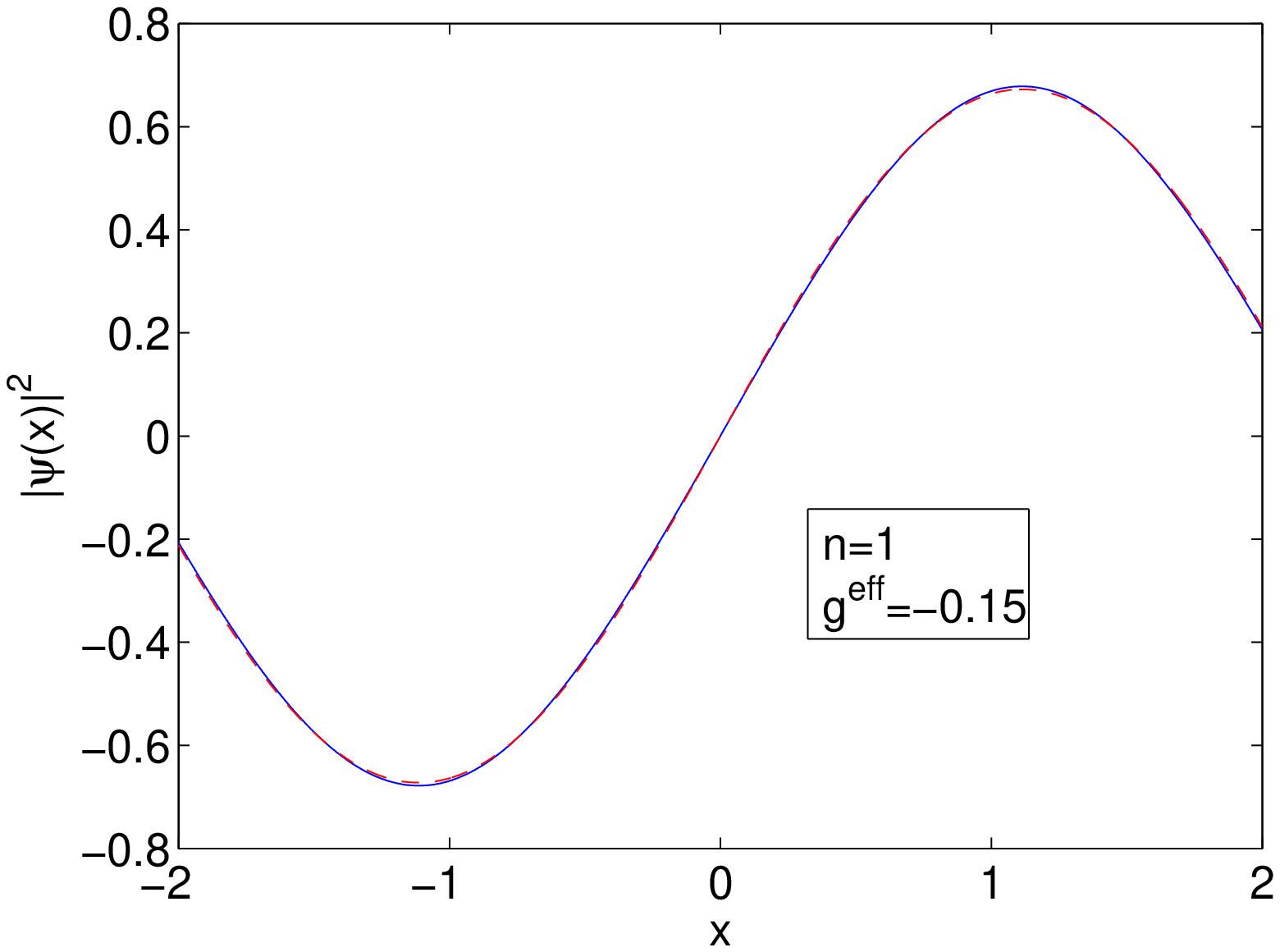}
\caption{\label{fig-WF} {(Color online) Wavefunctions of the two lowest bound states of the potential $V_0=-10$, $a=2$ for  $g^{\rm eff}=-0.1$ (\textcolor{blau}{'-'}) in comparison with the respective wavefunctions of the linear system (\textcolor{rot}{'- -'}). The normalization $\int_{-a}^a |\psi(x)|^2 \rd x =1$ has been used.}}
\end{figure}

\begin{figure}[htb]
\centering
\includegraphics[width=8cm,  angle=0]{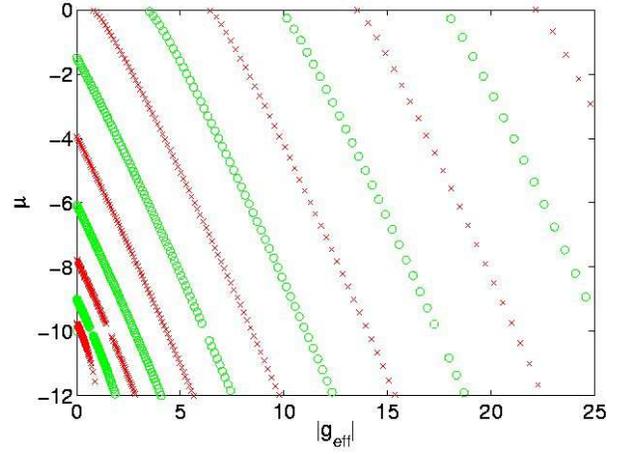}

\caption{\label{fig-full} {(Color online)
Shift of the bound states towards lower chemical potentials in dependence of $|g^{\rm eff}|$ for $p_\pm \le 0.99$ for the potential $V_0=-10$, $a=2$, for
symmetric (\textcolor{rot}{$\times$}),
and antisymmetric (\textcolor{gruen}{$\circ$}) bound states (attractive nonlinearity).
 }}
\end{figure}

Figure \ref{fig-WFT1} shows the squared modulus of wavefunctions at the transition from resonances to bound states. As in the case of resonances the functions are symmetric and have $n$ minima (cf. Fig.~\ref{fig-WFQ-}). The quantum number $n$ counts all the resonances and bound states defined above and is equal to the number of minima of the squared modulus of the wavefunction. For bound states, these minima are also nodes of the wavefunction. Thus, the famous oscillation theorem of linear quantum mechanics (cf. \cite{Land77}) is also valid in the nonlinear case considered here.

In order to study the shift of the bound state energies due to mean-field interaction it is useful to introduce an effective nonlinear interaction strength
\be
   g_\pm^{\rm eff}=\frac{g}{2a}\int_{-a}^a|\psi_\pm(x)|^2 dx
   \label{Rechteck_geff}
\ee
(cf. \cite{Witt05}) in the interaction region $|x| \le a$.
The integration can be carried out analytically. We obtain
\begin{eqnarray}
   g_-^{\rm eff}=&-&\frac{\hbar^2 \varrho_-}{ma} [E(\varrho_- \, a+K(p_-)|p_-)-E(p_-) \nonumber \\
                 &-&(1-p_-)\varrho_- \, a]
   \label{gef-}
\end{eqnarray}
for antisymmetric and
\be
   g_+^{\rm eff}=-\frac{\hbar^2 \varrho_+}{ma} \left[E(\varrho_+ \, a|p_+)-(1-p_+)\varrho_+ \, a
   \right]
   \label{gef+}
\ee
for symmetric bound states where $E(p)$ and $E(u|p)$ are the complete and incomplete elliptic integral of the second kind, respectively. The critical effective nonlinear interaction strength for the transition from resonances to bound states is given by
\be
   g_{T_\pm}^{\rm eff}=\frac{g_{T_\pm}|A|^2}{p_{T_\pm}} \left[ \frac{E(p_{T_\pm})}{K(p_{T_\pm})}-(1-p_{T_\pm}) \right] .
\ee

As seen from Eqs. (\ref{gef-}) and (\ref{gef+}), $|g^{\rm eff}(p_\pm)|$ increases faster for bigger wavenumbers $\varrho_\pm$ (respectively bigger chemical potentials $\mu$, cf. Fig.~\ref{fig-geff(p)}). This means that for a given value of $g^{\rm eff}$,  states which are strongly bound are more affected by the nonlinear interaction than weakly bound states. As in the case of resonances this is due to the fact that for weakly bound states the energy of the mean-field interaction is much smaller than the kinetic energy and thus has a comparatively weak influence. This can also be observed in Fig.~\ref{fig-WF} which compares the wavefunctions of the two lowest bound stats of the potential $V_0=-10$, $a=2$ for $g^{\rm eff}=-0.15$ with the respective wavefunctions of the linear system ($g^{\rm eff}=0$). The wavefunction of the state $n=0$ differs much more clearly from its linear counterpart than the wavefunction of the state $n=1$.

\begin{figure}[htb]
\centering
\includegraphics[width=8cm,  angle=0]{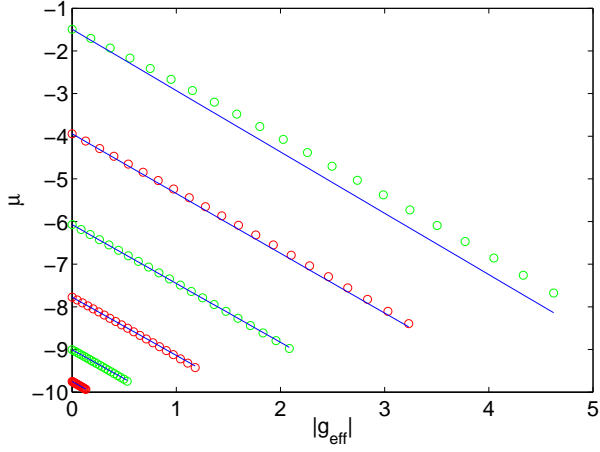}

\caption{\label{fig-app} {(Color online)
Shift of the bound states towards lower chemical potentials in dependence of $|g^{\rm eff}|$ for $p_\pm \le 0.4$ for the potential $V_0=-10$, $a=2$.
'$\circ$': numerically exact values 
\textcolor{blau}{'-'}: respective linear approximation due to Eq.~(\ref{mu_geff_app}) (attractive nonlinearity). The sudden disappearance of the curves at $g^{\rm eff}(p=0.4)$ is due to the fact that in this figure
the Jacobi elliptic parameter was ''artificially'' limited to
$p \le 0.4$ in order to avoid confusion and for a better comparison of the exact values and the approximation.
 } }
\end{figure}

Figure \ref{fig-full} shows the shift of bound states in dependence of $|g^{\rm eff}|$. The appearance of new bound states can be observed. For large values of $|g^{\rm eff}|$, which correspond to elliptic parameters $p>0.5$, the states all lie below the potential well depth $V_0=-10.$
For small values of $p_\pm$ the chemical potential $\mu(g_\pm^{\rm eff})$ can be approximated linearly in terms of $g_\pm^{\rm eff}$,
\begin{eqnarray}
   g_-^{\rm eff}& = &-\frac{\hbar^2 \varrho_-^2 p_-}{2ma}\int_{-a}^{a}\cn^2(\varrho_- \, x+K(p_-)) \rd x \\ \nonumber
            & = & -p_-(E_0-V_0)\left( 1-\frac{\sin(2qa)}{2qa} \right)+\mathcal{O}(p_-^2) \, ,
 \label{gfeld}
\end{eqnarray}
where
\be
   E_0=V_0+\hbar^2q^2/(2m)
   \label{E0}
\ee
is the energy of the respective bound state of the linear system. For symmetric states
we get
\be
   g_+^{\rm eff} =  -p_+(E_0-V_0)\left( 1+\frac{\sin(2qa)}{2qa} \right)+\mathcal{O}(p_+^2) .
\ee
Then, for small values of $p_\pm$, we have
\be
   p_\pm \approx - g_\pm^{\rm eff} \left[ (E_0-V_0)\left( 1 \pm \frac{\sin(2qa)}{2qa} \right) \right]^{-1}
   \label{p_pm_app}
\ee
and the chemical potential is given by
\be
   \mu_\pm=V_0+\frac{\hbar^2 \varrho_\pm^2}{2m}(1-2p_\pm) \, ,
\ee
where $\varrho_\pm$ is the wavenumber of a Jacobi elliptic function $\cn$ with quarter period $K(p_\pm)$. In the limit $p_\pm \rightarrow 0$ the $\cn$-function becomes a trigonometric function with quarter period $\pi/2$ and wavenumber $q$. This motivates the approximation $\varrho_\pm \approx 2 q K(p_\pm)/\pi$ which leads to
\be
   \mu_\pm \approx \frac{\hbar^2 q^2}{2m} \frac{4}{\pi^2}(1-p_\pm)K^2(p_\pm) .
\ee
Inserting the equations (\ref{E0}), (\ref{p_pm_app}) and expanding $K^2(p)(1-2p)=(\pi/2)^2(1-\frac{3}{2}p)+\mathcal{O}(p^2)$ we obtain
\be
   \mu_\pm \approx E_0+\frac{3}{2} \left(1\pm \frac{\sin(2qa)}{2qa} \right)^{-1}g_\pm^{\rm eff} .
   \label{mu_geff_app}
\ee
A similar relation was derived for a delta-shell potential \cite{Witt05}.
For small values of $p_\pm$ this approximation is in good agreement with the numerically exact calculations (cf. fig \ref{fig-app}).
In the limit $V_0 \rightarrow -\infty$ and thus $q \rightarrow \infty$ we obtain the formula
\be
   \mu_\pm \approx E_0 + \frac{3}{2} g_\pm^{\rm eff}
\ee
for an infinitely deep square well potential \cite{Carr00a,Carr00b,Dago00}.

\section{Repulsive nonlinearity: bound states}
\label{Rep_bs}

As in the previous section we look for solutions of odd or even parity.
We make the ansatz

\be
 \psi(x)_+=  \left\{ \begin{array}{cc}
                           \gamma_+ \, \re^{\kappa x}   &    x<-a \\
                 I_+ \, \mbox{sn}(\varrho_+ \, x +K(p_+) |p_+)     & |x| \le a\\
                           \gamma_+ \, \re^{-\kappa x}  &     x>a \\
                    \end{array}
              \right.
\ee
for even solutions and
\be
 \psi(x)_-=  \left\{ \begin{array}{cc}
                                  \gamma_- \re^{\kappa x}   &    x<-a \\
                 I_- \mbox{sn}(\varrho_- x |p_-)     &  |x| \le a\\
                                 -\gamma_- \re^{-\kappa x}  &     x>a \\
                    \end{array}
              \right.
\ee
for odd solutions where $\kappa=\sqrt{-2m \mu/\hbar}$.

For $|x| \le a$ the solutions are given by Jacobi elliptic $\sn$-functions with amplitude
\be
    I_\pm=\sqrt{\frac{\hbar^2 \varrho_\pm^2 p_\pm}{g m}}
    \label{I_pm_2}
\ee
and wavenumber
\be
   \varrho_\pm =\sqrt{\frac{2m(\mu-V)}{\hbar^2(1+p_\pm)}}.
   \label{rho_pm_2}
\ee

Again the wavefunctions and their derivatives must be continuous at $x=\pm a$. Thus we obtain the equations
   \be
      \kappa \ \mbox{sn}(u|p_+) +\varrho_+ \, \mbox{cn}(u|p_+)
      \mbox{dn}(u|p_+) = 0
     \label{tan_nl_2}
   \ee
for even parity, with $u=\varrho_+ \, a+K(p_+)$, and
  \be
      \kappa \, \mbox{sn}(\varrho_- \, a|p_-) +\varrho_- \, \mbox{cn}(\varrho_- \, a|p_-) \mbox{dn}(\varrho_- \, a|p_-) = 0
      \label{cot_nl_2}
   \ee
for odd parity, which determine the bound states of the system. In the limit $p \rightarrow 0$ the conditions (\ref{tan_nl_2}) and (\ref{cot_nl_2}) converge to the equations (\ref{tanqa}) and (\ref{cotqa}) of the linear system.

For the case of an attractive nonlinearity, we have shown that resonances with even/odd quantum number $n$ become bound states of even/odd parity due to the attractive mean-field interaction. In a similar way one can show for the case of repulsive nonlinearity that even/odd bound states become resonances with even/odd quantum number $n$.

As before, we use the normalization $|\gamma_\pm|=|A|$. Thus we obtain the conditions
\be
   \varrho_\pm a=n K(p_\pm)
\ee
where
\be
     n= \left\{ \begin{array}{cc}
                                  2 j_+, \, \, j_+ \in \mathbb{N}_0   &    \text{even parity} \\
                                  2 j_-+1, \, \, j_- \in \mathbb{N}_0   &    \text{odd parity} \, .
                     \end{array}
              \right.
\ee
The transition occurs at
\be
   g_{T\pm} =\frac{\hbar^2n^2K(p_{T\pm})^2 p_{T\pm}}{m l^2 |A|^2} \, ,
\ee
where the transition parameters $p_\pm$ are given by
\be
   (1+p_{T\pm})K^2(p_{T\pm})=\frac{2m|V_0|^2a^2}{\hbar^2 n^2}
\ee
or
\be
   p_{T\pm}=-\frac{24}{25}+\sqrt{\left(\frac{24}{25}\right)^2-\zeta(n)}+\mathcal{O}(p_{T\pm})^3
\ee
with
\be
   \zeta(n)=\frac{32}{25}\left(1-\frac{8m|V_0|a^2}{\hbar^2\pi^2n^2} \right) .
\ee
Since $p_{T_\pm}$ has to be positive this implies the conditions
\be
   j_\pm \le N_\pm \, ,
\ee
where $N_+$ and $N_-$ are the numbers of even and odd bound states of the linear system. This expresses the obvious fact that the number of newly created resonances cannot be greater than the number of bound states of the linear system.

As in the case of an attractive nonlinearity, the quantum number $n$ numbers the the resonances as well as the symmetric and antisymmetric bound states and is equal to the number of the minima of the squared modulus of the wavefunction. In the case of bound states, these minima are also zeros of the wavefunction, so that the oscillation theorem is valid.

\begin{figure}[htb]
\centering
\includegraphics[width=8cm,  angle=0]{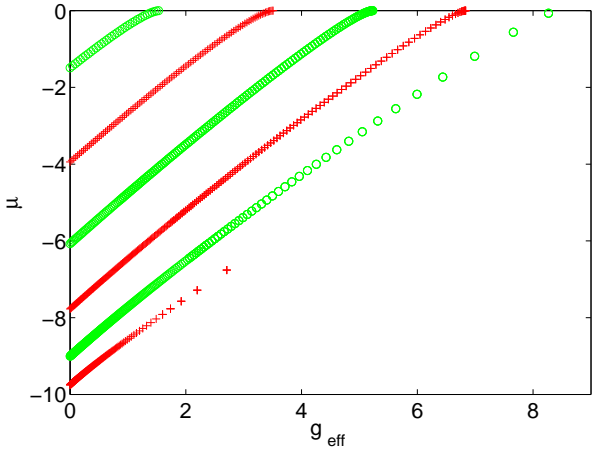}

\caption{\label{fig-fullR} {(Color online)
Shift of the bound states towards lower chemical potentials in dependence of $g^{\rm eff}$ for $p_\pm \le 0.995$ for the potential $V_0=-10$, $a=2$.
\textcolor{rot}{'+'}: symmetric bound states,
\textcolor{gruen}{'$\circ$'}: antisymmetric bound states. The sudden disappearance of the curves at
$g^{\rm eff}(p=0.995)$ is due to the fact that in this figure the
Jacobi elliptic parameter had to be ''artificially'' limited to $p \le 0.995$ for numerical reasons.
 }}
\end{figure}

In terms of the effective nonlinearity defined by Eq.~(\ref{Rechteck_geff}), we obtain
\be
   g_-^{\rm eff}=\frac{\hbar^2 \varrho_-}{ma} \left[\varrho_- \, a-E(\varrho_- \, a|p_-) \right]
\ee
for antisymmetric bound states and
\be
   g_+^{\rm eff}=\frac{\hbar^2 \varrho_+}{ma} \left[ \varrho_+ \, a + E(p_+) -E(\varrho_+ \, a+K(p_+)|p_+)
   \right]\,
\ee
for symmetric bound states.
For the transition we get in both cases
\be
   g_{T_\pm}^{\rm eff}=\frac{g_{T_\pm}|A|^2}{p_{T_\pm}} \left[1- \frac{E(p_{T_\pm})}{K(p_{T_\pm})} \right] .
\ee
The magnitude of $g^{\rm eff}$ has the same qualitative dependence on $p_\pm$ as in the attractive case so that the nonlinearity has a stronger effect on lower bound states. 

Figure \ref{fig-fullR} shows how the bound states of the system are shifted towards higher chemical potentials if $g^{\rm eff}$ is increased, until they reach $\mu=0$ and develop into resonance states.
Approximating the chemical potential linearly in terms of $g^{\rm eff}$ results in
\be
   \mu_\pm \approx E_0+\frac{3}{2} \left(1\pm \frac{\sin(2qa)}{2qa} \right)^{-1}g^{\rm eff} \,
   \label{mu_geff_app_2}
\ee
which is identical with the respective Eq.~(\ref{mu_geff_app}) for attractive nonlinearities.

\section{Dynamics}
\label{dyn}

\begin{figure}[htb]
\centering
\includegraphics[width=8cm,  angle=0]{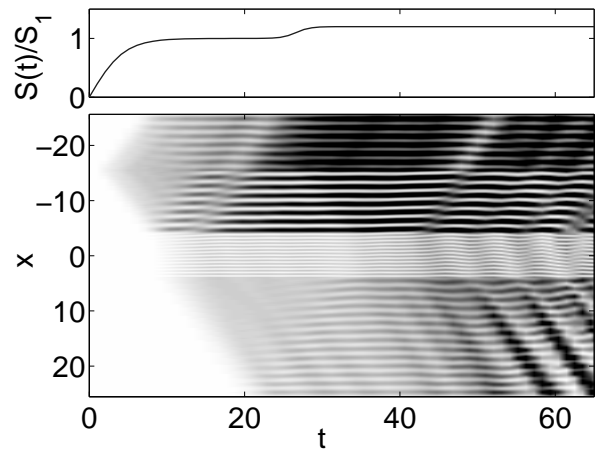}
\caption{\label{fig-trans_beak1} {
Instability due to a beak-to-beak crossing scenario in the transmission through
a square-well potential with $V_0 = -12$ and.
Shown is the squared modulus of the wave function $|\psi(x,t)|^2$ for $\mu=2$ in a grayscale
plot (lower panel) and the source strength $S(t)$ (upper panel) for comparison (repulsive nonlinearity).
}}
\end{figure}

In this section we briefly discuss the implications of the results for the nonlinear eigenstates
presented above on the dynamics. To this end we assume that a source
outside of the square-well potential emits condensed atoms with a
fixed chemical potential $\mu$.
In fact, we numerically solve the {\it time-dependent} NLSE with an additional
delta-localized source term (cf. \cite{Paul05})
\begin{eqnarray}
  \ri \frac{\partial \psi(x,t)}{\partial t} &=&
  \left( - \frac{1}{2} \frac{\partial^2}{\partial x^2}
  + V(x) + g(x) |\psi(x,t)|^2 \right) \psi(x,t) \nonumber \\
  &+& S(t) \exp(-\ri \mu t ) \delta(x-x_0) \, ,
  \label{eqn-tdnlse-source}
\end{eqnarray}
where units with $\hbar=m=1$ are used.
As above, $V(x) = V_0 H(a - |x|)$ is a square-well potential with depth $V_0 = -12$ and half-width $a=4$. The nonlinearity is non-zero only within the square-well $g(x) = gH (a - |x|)$. Here, $H(x)$ denotes
the Heaviside step function. The source is located at $x_0=-15.5$.
The initial state is an empty waveguide $\psi(x,t=0) = 0$. The source strength
$S(t)$ is then ramped up adiabatically, so that finally a steady transmitting
solution $\psi(x,t_{\rm end})$ is found.
To determine the incident current and the transmission coefficient, a
superposition of plane waves with momenta $k = \pm \sqrt{2\mu}$ is fitted
to the wave function $\psi(x,t_{\rm end})$ for $x_0 < x < -a$ (upstream region)
and $x > a$ (downstream region).
The NLSE is numerically integrated in real space using a predictor-corrector
step \cite{Cerb98}.
To avoid reflections at the edges of the computational interval, one-way boundary
conditions are used as described in \cite{Shib91}.
An example of the wave function calculated in this way is shown in Fig.~\ref{fig-trans_beak1}.

\begin{figure}[htb]
\centering
\includegraphics[width=8cm,  angle=0]{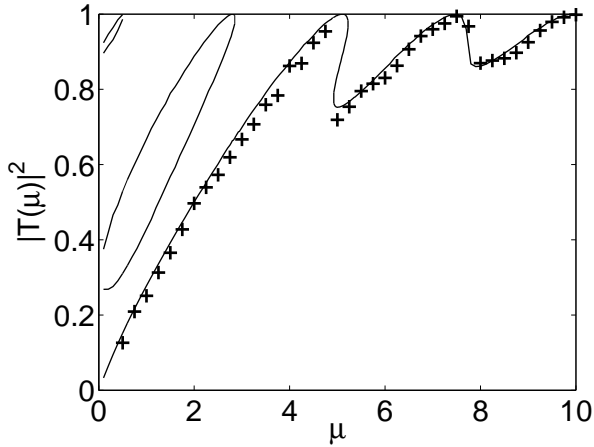}
\caption{\label{fig-trans_num_g=4} {
Transmission of the square-well potential $V_0 = -12, a = 4$ with a repulsive nonlinearity
$g|A|^2 =4$. Shown is the transmission coefficient of the stationary states calculated
as described in the previous sections (solid lines) in comparison to the transmission
coefficients obtained from a numerical solution of the time-dependent NLSE as described in the text (crosses).
}}
\end{figure}

The results of these simulations are shown in Fig.~\ref{fig-trans_num_g=4}, where the
transmission coefficient is plotted in dependence of the chemical potential $\mu$ for a
fixed repulsive nonlinearity $g|A|^2 =4$.
Firstly, we notice that the final state $\psi(x,t_{\rm end})$ is indeed a stationary
solution as discussed in the previous sections. The transmission coefficient
$|T(\mu)|^2$ follows closely one of the steady state results which are computed as
described in the previous sections.
The small deviations are caused by numerical errors, in fact mainly by the (still not
perfectly absorbing) boundary conditions.
Secondly, the final state $\psi(x,t_{\rm end})$ is the one with the smallest
transmission.
This behaviour may suppress resonant transport, because a resonance state is not populated
if different stationary states exist with a smaller transmission. This can be seen in
Fig.~\ref{fig-trans_num_g=4} around $\mu \approx 5$ and $\mu \lesssim 2$.
However it is possible to access states with a higher transmission by a suitable
preparation procedure as it has been shown in \cite{Paul05}.
If there is no bistability in the vicinity of a resonance (e.g. around $\mu \approx 7.5$
in Fig.~\ref{fig-trans_num_g=4}), resonant transport is found.

A more complicated situation arises when a beak-to-beak bifurcation scenario occurs
(cf. Fig.~\ref{fig-T_g+}). As an example, we consider the solution of
the time-dependent NLSE with a source term as in (\ref{eqn-tdnlse-source}) for
the same square-well potential as above ($V_0 = -12$, $a=4$) and a fixed
chemical potential $\mu = 2$.
Again the source strength is increased adiabatically \cite{Liu03} to a level $S(t) = S_1$ so
that the system assumes a stationary transporting state, where $g|A|^2 = 3.5$ is
well below the critical value for the beak-to-beak bifurcation. Then the source
strength is again slowly increased up to $S(t) = S_2$, so that $g|A|^2$
becomes larger than the critical value of the beak-to-beak bifurcation.
The lowly-transmitting stationary state ceases to exist and the time evolution
cannot follow the stationary solutions adiabatically any longer.
This is shown in Fig.~\ref{fig-trans_beak1}.
Indeed one observes that the system becomes unstable and finally does not assume
a stationary state any more.
However, the transmission through the square-well increases.
The beak-to-beak crossing scenario and its implications on the dynamics will be
discussed in detail in a subsequent paper \cite{Grae05}.

\section{Conclusion}

The scattering of a  BEC by a finite square well potential has been discussed in terms of stationary states of the nonlinear Schr\"odinger equation. Neglecting the mean-field interaction outside the potential ingoing and outgoing waves and thus amplitudes of reflection and transmission can be defined.
The transmitted flux shows a bistable behaviour in the vicinity of the resonances. For repulsive nonlinearities, the transmission coefficient also shows beak-to-beak bifurcations. Wherever these beak-to-beak scenarios occur, the scattering phase shows loops.

An analytical expression for the position of the resonances is derived. The transitions from resonances to bound states and vice versa due to attractive/repulsive mean-field interaction are proven analytically and an explicit formula for the transition interaction strength $g_T$ is found. Furthermore, the positions of the bound states of the system are calculated and compared with the case of vanishing interaction $g=0$. It is found that the oscillation theorem also holds for interaction parameters $g \ne 0$.

Finally, we solve the time-dependent NLSE numerically in order to discuss the dynamics of the system for $g>0$. If the amplitude $|A|$ of the incoming flux is slowly ramped up for a fixed value of the chemical potential $\mu$, the system populates the lowest branch of the stationary transmission coefficient, as it was first shown in \cite{Paul05} for a double barrier potential, so that resonant transport is usually be suppressed. In the vicinity of a beak-to beak bifurcation, the system can become unstable if $g|A|^2$ becomes larger than the critical value of the beak-to-beak bifurcation so that the lowest branch of the stationary transmission coefficient suddenly ceases to exist. These beak-to-beak crossing scenarios will be further investigated in a subsequent paper \cite{Grae05}.

\section*{Acknowledgements}
Support from the Deutsche Forschungsgemeinschaft
via the Graduiertenkolleg ``Nichtlineare Optik und Ultrakurzzeitphysik''
is gratefully acknowledged. We thank Peter Schlagheck and Tobias Paul for inspiring discussions.


\bibliography{abbrev,dipldiss,paper60,paper70,paper80,paper90,paper00,publko,rest,mybib}

\end{document}